\documentclass[10pt, final, twocolumn, twoside, romanappendices]{IEEEtran}

\IEEEoverridecommandlockouts

\usepackage{graphicx,cite,acronym,amsmath,amssymb,amsfonts,color,subfigure,floatflt,stfloats,bm,textgreek,multirow}
\usepackage{epsfig,epstopdf,psfrag}
\usepackage[ruled,vlined]{algorithm2e}
\SetKwInput{KwInput}{Input}                
\SetKwInput{KwOutput}{Output}  

\usepackage[table]{xcolor}
\usepackage{bm}

\usepackage[cmintegrals]{newtxmath}
\usepackage{graphics,hhline,array,cite,bbm}

\usepackage{latexsym}
\usepackage{theorem}
\usepackage{times}
\usepackage{makeidx}
\DeclareGraphicsExtensions{.png,.jpg,.eps}

\acrodef{ESPAR}{electrically steerable passive array radiator}

\acrodef{SIM}{stacked intelligent metasurface}

\acrodef{SE}{spectral efficiency}

\acrodef{DAC}{digital-to-analog converter}

\acrodef{DMA}{dynamic metasurface antenna}

\acrodef{SINR}{signal-to-interference noise ratio}

\acrodef{ESIT}{electromagnetic signal and information theory} 

\acrodef{ELAA}{extremely large antenna arrays} 

\acrodef{DSA}{dynamic scattering array}

\acrodef{ULA}{uniform linear array}

\acrodef{UCA}{uniform circolar array}

\acrodef{IIoT}{industrial Internet-of-things}

\acrodef{IT}{information theory}

\acrodef{SRE}{smart radio environment}

\acrodef{EMO}{electromagnetic object}

\acrodef{SVD}{singular value decomposition}

\acrodef{PSWF}{prolate spheroidal wave function}

\acrodef{CR}{channel response}

\acrodef{BS}{base station}

\acrodef{MS}{mobile station}

\acrodef{UE}{user equipment}

\acrodef{MIMO}{multiple-input multiple-output}

\acrodef{MISO}{multiple-input single-output}

\acrodef{RIS}{reconfigurable intelligent surface}

\acrodef{IRS}{intelligent reconfigurable surface}

\acrodef{LIS}{large intelligent surface}

\acrodef{MIS}{medium intelligent surface}

\acrodef{SIS}{small intelligent surface}

\acrodef{DoF}{degrees-of-freedom}

\acrodef{AF}{amplify \& forward}

\acrodef{DF}{detect \& forward}

\acrodef{JF}{just forward}

\acrodef{CSI}{channel state information}

\acrodef{RV}{random variable}

\acrodef{i.i.d.}{independent, identically distributed}

\acrodef{PSD}{power spectral density}

\acrodef{PDF}{probability distribution function}

\acrodef{CDF}{cumulative distribution function}

\acrodef{ch.f.}{characteristic function}

\acrodef{AWGN}{additive white Gaussian noise}

\acrodef{RSSI}{received signal strength indicator}

\acrodef{SNR}{signal-to-noise ratio}

\acrodef{LRT}{likelihood ratio test}

\acrodef{GLRT}{generalized likelihood ratio test}

\acrodef{GML}{generalized maximum likelihood}

\acrodef{LOS}{line-of-sight}

\acrodef{NLOS}{non-line-of-sight}

\acrodef{GDOP}{geometric dilution of precision}

\acrodef{GPS}{Global Positioning System}

\acrodef{FIM}{Fisher information matrix}

\acrodef{PEB}{position error bound}

\acrodef{WSN}{Wireless Sensor Network}

\acrodef{MAC}{medium access control}

\acrodef{RSS}{received signal strength}

\acrodef{RTT}{round-trip time}

\acrodef{MIMO}{multiple-input multiple-output}

\acrodef{MF}{matched filter}

\acrodef{ED}{energy detector}

\acrodef{ML}{maximum likelihood}

\acrodef{NL}{nonlinear}

\acrodef{MSE}{mean square error}

\acrodef{RMSE}{root mean square error}

\acrodef{ppm}{part-per-million}

\acrodef{PRP}{pulse repetition period}

\acrodef{ACK}{acknowledge}

\acrodef{UWB}{ultrawide bandwidth}

\acrodef{TNR}{threshold-to-noise ratio}

\acrodef{LOS}{line-of-sight}

\acrodef{LS}{least squares}

\acrodef{IR-UWB}{impulse radio UWB}

\acrodef{FCC}{Federal Communications Commission}

\acrodef{TH}{time-hopping}

\acrodef{PPM}{pulse position modulation}

\acrodef{PAM}{pulse amplitude modulation}

\acrodef{MUI}{multi-user interference}

\acrodef{PDP}{power delay profile}

\acrodef{PPP}{Poisson point process}

\acrodef{DS}{delay spread}

\acrodef{CED}{channel excess delay}

\acrodef{BPZF}{band-pass zonal filter}

\acrodef{SIR}{signal-to-interference ratio}

\acrodef{RFID}{radio frequency identification}

\acrodef{WPAN}{wireless personal area networks}

\acrodef{WWLB}{Weiss-Weinstein lower bound}

\acrodef{DP}{direct path}

\acrodef{MF}{matched filter}

\acrodef{MMSE}{minimum-mean-square-error}

\acrodef{SBS}{serial backward search}

\acrodef{NBI}{narrowband interference}

\acrodef{WBI}{wideband interference}

\acrodef{INR}{interference-to-noise ratio}

\acrodef{CIR}{channel impulse response}

\acrodef{ISI}{inter-symbol interference}

\acrodef{CPR}{channel pulse response}

\acrodef{LRT}{likelihood ratio test}

\acrodef{MUI}{multi-user interference}

\acrodef{EM}{electromagnetic}

\acrodef{CW}{continuous wave}

\acrodef{RF}{radiofrequency}

\acrodef{FCC}{Federal Communications Commission}

\acrodef{EIRP}{effective radiated isotropic power}

\acrodef{RCS}{radar cross section}

\acrodef{BAV}{balanced antipodal Vivaldi}

\acrodef{PRake}{partial Rake}

\acrodef{RTLS}{real time locating system}

\acrodef{CRB}{Cram\'{e}r-Rao bound}

\acrodef{ZZB}{Ziv-Zakai bound}

\acrodef{TOA}{time-of-arrival}

\acrodef{TOF}{time-of-flight}

\acrodef{WSN}{wireless sensor network}

\acrodef{MAC}{medium access control}

\acrodef{RSS}{received signal strength}

\acrodef{TDOA}{time difference-of-arrival}

\acrodef{RF}{radiofrequency}

\acrodef{RTT}{round-trip time}

\acrodef{AOA}{angle-of-arrival}

\acrodef{MF}{matched filter}

\acrodef{ED}{energy detector}

\acrodef{ML}{maximum likelihood}

\acrodef{MUR}{Multistatic radar}

\acrodef{HDSA}{high-definition situation-aware}

\acrodef{RRC}{root raised cosine}

\acrodef{OFDM}{orthogonal frequency division multiplexing}

\acrodef{IF}{intermediate frequency}

\acrodef{PHY}{physical layer}

\acrodef{S-V}{Saleh-Valenzuela}

\acrodef{UHF}{ultra-high frequency}

\acrodef{PR}{pseudo-random}

\acrodef{SoC}{System on Chip}

\acrodef{SoP}{System on Package}

\acrodef{SPMF}{Single-Path Matched Filter}

\acrodef{IMF}{Ideal Matched Filter}

\acrodef{SCR}{signal-to-clutter ratio}

\acrodef{BEP}{bit error probability}

\acrodef{BER}{bit error rate}

\acrodef{WSR}{wireless sensor radar}

\acrodef{HPBW}{half power beam width}

\acrodef{LEO}{localization error outage}

\acrodef{WSS}{wide-sense stationary}

\acrodef{TR}{time-reversal}

\acrodef{WSSUS}{WSS with uncorrelated scattering}

\acrodef{GP}{Gaussian process}

\acrodef{IMU}{inertial measurement unit}

\newcommand{\Real}[1]{\Re \left \{ #1\right \}}
\newcommand{\Imag}[1]{\Im \left \{ #1\right \}}

\newcommand{\EX}[1] {{\mathbb{E}}\left\{{#1}\right\}}

\newcommand{\boldA} {{\bf{A}}}

\newcommand{\boldV} {{\bf{V}}}

\newcommand{\boldt} {{\bf{t}}}
\newcommand{\boldU} {{\bf{U}}}

\newcommand{\boldH} {{\bf{H}}}

\newcommand{\boldB} {{\bf{B}}}

\newcommand{\boldQ} {{\bf{Q}}}

\newcommand{\boldI} {{\bf{I}}}

\newcommand{\boldn} {{\bf{n}}}
\newcommand{\boldx} {{\bf{x}}}
\newcommand{\boldy} {{\bf{y}}}

\newcommand{\diag}[1]{{\rm diag} \left ( #1 \right )}

\newcommand{\argmin}[1]{\underset{{#1}}{\operatorname{arg \, min}}}
\newcommand{\minimize}[1]{\underset{{#1}}{\operatorname{minimize}}}

\newcommand{\ctranspose}{^{\mathsf{H}}}
\newcommand{\transpose}{^{\mathsf{T}}}

\newcommand{\Na} {N_{\text{A}}}
\newcommand{\Ns} {N_{\text{S}}}

\newcommand{\Ptx} {P_{\text{T}}}
\newcommand{\Prx} {P_{\text{R}}}
\newcommand{\Prad} {P_{\text{rad}}}
\newcommand{\Preact} {P_{\text{react}}}

\newcommand{\Pa} {P_{\text{a}}}

\newcommand{\Pd} {P_{\text{d}}}

\newcommand{\etam} {\eta_{\text{m}}}
\newcommand{\etad} {\eta_{\text{d}}}

\newcommand{\Gr} {G_{\text{R}}}

\newcommand{\bi} {\mathbf{i}}
\newcommand{\bit} {\bi_{\text{T}}}
\newcommand{\bia} {\bi_{\text{A}}}
\newcommand{\bis} {\bi_{\text{S}}}

\newcommand{\bv} {\mathbf{v}}
\newcommand{\bvt} {\bv_{\text{T}}}
\newcommand{\vg} {{v_{\text{G}}}}
\newcommand{\bvg} {\bv_{\text{G}}}
\newcommand{\bva} {\bv_{\text{A}}}

\newcommand{\bvs} {\bv_{\text{S}}}

\newcommand{\btheta} {\boldsymbol{\theta}}
\newcommand{\balpha} {\boldsymbol{\alpha}}
\newcommand{\bphi} {\boldsymbol{\phi}}
\newcommand{\bpsi} {\boldsymbol{\psi}}

\newcommand{\bPhi} {\boldsymbol{\Phi}}

\newcommand{\bLambda} {\boldsymbol{\Lambda}}

\newcommand{\bZ} {\mathbf{Z}(f)}
\newcommand{\bZa} {\mathbf{Z}_{\text{A}}(f; \btheta)}
\newcommand{\bZaa} {\mathbf{Z}_{\text{AA}}(f)}
\newcommand{\bZas} {\mathbf{Z}_{\text{AS}}(f)}
\newcommand{\bZsa} {\mathbf{Z}_{\text{SA}}(f)}
\newcommand{\bZss} {\mathbf{Z}_{\text{SS}}(f)}
\newcommand{\bZs} {\mathbf{Z}_{\text{S}}(f; \btheta)}

\newcommand{\bZl} {\mathbf{Z}_{\text{L}}(f; \bphi)}

\newcommand{\bWdo} {\mathbf{W}_{\text{D}}}
\newcommand{\bWdk} {{\mathbf{W}_{\text{D}_k}}}

\newcommand{\bWdhatk} {{\hat{\mathbf{W}}_{\text{D}_k}}}

\newcommand{\bWdd}[1] {\mathbf{W}_{\text{D}_{#1}}}
\newcommand{\Wd}[1] {W_{\text{D}_{#1}}}

\newcommand{\bWem}[1] {\mathbf{W}_{\text{EM}}  \left({#1}; \bpsi \right )}
\newcommand{\bWemhat}[1] {\mathbf{W}_{\text{EM}}   \left ({#1}; \hat{\bpsi}  \right )}

\newcommand{\bHc}[1] {\mathbf{H}_{\text{c}}({#1})}
\newcommand{\bhc}[1] {{\mathbf{h}_{\text{c}}}({#1})}

\newcommand{\bHoptk} {\mathbf{H}^{(\text{opt})}_k}
\newcommand{\bHopt} {\mathbf{H}^{(\text{opt})}}

\newcommand{\bZm}[1] {\mathbf{Z}_{\text{M}}({#1};\btheta)}

\newcommand{\bYm}[1] {\mathbf{Y}_{\text{M}}({#1};\bphi)}

\newcommand{\boldzero} {\boldsymbol{0}}

\begin{document}
\title{Over-the-air Multifunctional Wideband Electromagnetic Signal Processing\\ using Dynamic Scattering Arrays}

\author{
\IEEEauthorblockN{Davide~Dardari,~\IEEEmembership{Fellow,~IEEE}}
\IEEEcompsocitemizethanks{\IEEEcompsocthanksitem 
 D.~Dardari is with the 
   Dipartimento di Ingegneria dell'Energia Elettrica e dell'Informazione ``Guglielmo Marconi"  (DEI), WiLAB-CNIT, 
   University of Bologna, Cesena Campus, 
   Cesena (FC), Italy, (e-mail: davide.dardari@unibo.it). 
    }
}

\maketitle

\begin{abstract}


To meet the stringent requirements of next-generation wireless networks, \ac{MIMO} technology is expected to become massive and pervasive. Unfortunately, this could pose scalability issues in terms of complexity, power consumption, cost, and processing latency. Therefore, novel technologies and design approaches, such as the recently introduced holographic \ac{MIMO} paradigm, must be investigated to make future networks sustainable. 
In this context, we investigate the concept of a \ac{DSA} as a versatile \ac{EM} structure capable of performing joint wave-based computing and radiation by moving the processing from the digital domain to the \ac{EM} domain. We provide a general, wideband analytical framework for modeling the \ac{DSA}, which includes a power matching network and realistic reconfigurable loads. Then we introduce specific design algorithms, and apply them to various use cases. We demonstrate that some recent \ac{EM} processing structures can be seen as particular cases of our general framework.  The examples presented in the numerical results corroborate the potential of \acp{DSA} to reduce complexity and the number of \ac{RF} chains in holographic \ac{MIMO} systems while achieving enhanced \ac{EM} wave processing and radiation flexibility for tasks such as beamforming and single- and multi-user \ac{MIMO}, also exhibiting superdirectivity capabilities.

\end{abstract}




\section{Introduction}


\IEEEPARstart{N}{ext}-generation wireless systems are expected to provide enhanced performance in terms of capacity, reduced latency, and new functionalities such as integrated sensing and communication \cite{DanAmiShiAlo:20,Pre:J24}. This trend is driving the investigation of fundamental limits using physically consistent models, new technologies, and novel design paradigms to approach them \cite{San:19,DarDec:J21,BjoEldLarLozPoo:23,YouCaiLiuDiRDumYen:25,CasYanChaHea:25}.

One direction of research is the utilization of high-frequency bands in the millimeter wave and THz ranges, coupled with the integration of \ac{ELAA} \cite{WanZhaDuShaAiNiyDeb:23}, which is expected to open up the potential to utilize the extra \ac{DoF} provided by the channel in the radiating near-field propagation region, even under \ac{LOS} conditions  \cite{PhaIvrGraCreTanNos:18, DecDar:J21,LiuXuWanMuHan:23, ZhaShlGuiDarImaEld:J22, ZhaShlGuiDarEld:J23, TorDecDar:J23}.  However, this is pushing current technology towards insurmountable barriers in hardware complexity and power consumption. This issue, also known as \emph{digital bottleneck}, poses serious challenges for the sustainability of future wireless networks. 
In \ac{MIMO}-based systems, hybrid digital-analog solutions have been extensively explored to partially alleviate these issues by reducing the number of \acf{RF} chains and the digital processing burden, albeit at the expense of flexibility \cite{AlkElALeuHea:14,CasYanChaHea:25}. 

A promising approach toward sustainability is to delegate part of the signal processing directly to the \ac{EM} level, known as \ac{ESIT} \cite{Dar:JS25,ZhuWanDaiDebPoo:22,DiRMig:24,BjoChaHeaMarMezSanRusCasJunDem:24}.
 This can be achieved by designing reconfigurable \ac{EM} environments \cite{Dar:J24} using novel \ac{EM} metamaterials devices to perform basic processing functions (e.g., spatial first derivative) \cite{Sil:14}, \acp{RIS}, or the recently introduced \acp{SIM} \cite{AnXuNgAleHuaYueHan:23}.  More in general, holographic \ac{MIMO} surfaces are envisioned as an efficient implementation of large antenna systems, advancing beyond massive \ac{MIMO} and \acp{ELAA}  using simplified and reconfigurable hardware that is reduced in size, cost and power consumption, and facilitate signal processing in the \ac{EM} domain \cite{GonGavJiHuaAleWeiZhaDebPooYue:24}. 

Within this context, an appealing recent solution to realize flexible and compact \ac{EM} processing structures is given by the \acf{DSA} \cite{Dar:C24,Dar:J24b}. A \ac{DSA} can be viewed as a generalization of reactively controlled arrays, initially introduced in \cite{Har:78}, and their evolution into \acp{ESPAR} \cite{KalKanPap:13} and \acp{SIM}, toward the realization of multifunctional wave processing.  A \ac{DSA} comprises a limited number of active antenna elements, each connected to an \ac{RF} chain (input), and is surrounded by a cluster of many reconfigurable passive scatterers. These scatterers interact within the reactive near field, enabling \ac{EM} processing and radiation to occur jointly ``over the air", thus offering several advantages with respect to other solutions, as investigated in this paper.

\subsection{Related State of the Art}

The idea of introducing passive scattering elements close to an antenna to exploit mutual coupling in the reactive near-field region and shape its radiation properties dates back to the early days of wireless transmissions. A pioneering work is that by H. Yagi published in 1928 that introduced the classic Yagi-Uda array antenna. This antenna consists of a set of linear parallel dipoles with a spacing of about $0.2-0.3 \lambda$, where the first dipole is active (i.e., fed with the signal) and the others act as passive scatterers \cite{Poz:97}.
Over the subsequent decades, this idea has been extensively developed in various directions within the antenna theory community. One milestone is given by the introduction of reactively controlled arrays, originally proposed by Harrington in \cite{Har:78}, where the passive scatterers can be reconfigured through the addition of programmable reactive loads to change the radiation characteristics of the antenna. Such an idea has evolved into \acp{ESPAR}, which has been widely investigated to realize programmable antennas, mainly using a single \ac{RF} chain and a ring of passive scatterers around it \cite{KalKanPap:13,AleVlaCon:14,BucJuaKamSib:20}. The typical design approach adopted for \acp{ESPAR} is the characteristic mode analysis which consists of designing the loads such that the currents flowing in the elements, needed to obtain the desired radiation diagram, represent the dominant scattering mode of the structure \cite{Har:78}.  Unfortunately, this simple approach is not accurate when applied to large structures in which secondary modes might not be negligible and hence deviate from the actual response from the desired one. Moreover, the approach cannot be applied in the case of multiple inputs, i.e., for multifunctional designs.    

In more recent times, the introduction of new technologies and materials has enabled the development of new antenna structures designed to meet the demands of higher performance, flexibility, frequency, and lower cost, as required by new-generation wireless systems.   
Among these advancements, metasurface antennas have received particular attention. They are composed of subwavelength elements printed on a grounded dielectric slab. These antennas exploit the interaction between a cylindrical surface wave, excited by a monopole, and an anisotropic impedance boundary condition realized with passive printed elements to produce an almost arbitrary aperture field \cite{FaeMinGonCamMarDelMac:19}. 
The dynamic counterpart to static metasurface antennas is the \ac{DMA}, a type of traveling wave antenna with reconfigurable subwavelength apertures. Since only one \ac{RF} chain is needed for each row of the planar array, \acp{DMA} offer a compelling solution, balancing flexibility with a reduction in \ac{RF} chains  \cite{ShlAleImaYonSmi:21,ZhaShlGuiDarEld:J23}. 
Static and dynamic metasurface antennas represent a promising avenue for the research aimed at reducing power consumption in wireless systems, thanks to their ability to shape signals at the \ac{EM} level. 
Other recent technologies include fluid and reconfigurable antennas  \cite{WonNewHaoTonCha:23,PriHarBlaKieKusFriPraMorSmi:04,RodCetJof:14}.

 An important step toward holographic \ac{MIMO} is the introduction of \acp{SIM} or stacked \ac{RIS} \cite{AnXuNgAleHuaYueHan:23,HasAnDiRDebYue:24,AnDiRDebPooYue:23,AnYueGuaDiRDebPooHan:24}.
The concept of \ac{SIM} originates from recent advances in optics and microwave circuits designed to perform machine learning tasks using layered \ac{EM} surfaces to mimic a deep neural network for image processing \cite{LiuMaLuo:22,ZeQiaXinWeiTie:24}. 
Similarly, a \ac{SIM} consists of a closed vacuum container with several stacked metasurface layers. The first layer is an active planar antenna array fed by a set of \ac{RF} chains. The \ac{EM} wave generated by the array travels through subsequent layers composed of numerous reconfigurable meta-atoms. The last layer is in charge of radiating the \ac{EM} field outward.
By properly configuring the transmission properties of each meta-atom in the \ac{SIM}, the system can manipulate the traveling \ac{EM} wave to produce a customized waveform shape. Examples of optimization techniques for \acp{SIM} to achieve various \ac{EM} wave processing functionalities can be found in \cite{AnYueGuaDiRDebPooHan:24,HasAnDiRDebYue:24,AnXuNgAleHuaYueHan:23,AnDiRDebPooYue:23}. Specifically, \cite{AnXuNgAleHuaYueHan:23} investigates a holographic  \ac{MIMO} link where both the transmitter and the receiver use a \ac{SIM}. The authors show that with a 7-layer \ac{SIM} with $\lambda/2$ meta-atom spacing, a good fit with the ideal \ac{SVD} precoding and combining tasks is obtained. In \cite{HasAnDiRDebYue:24} and \cite{AnDiRDebPooYue:23}, optimization schemes to realize custom radiation patterns and multi-user downlink beamforming are proposed, whereas the authors in \cite{AnYueGuaDiRDebPooHan:24} demonstrate the capability to perform a 2D-Fourier  transform in the wave domain.   
 Although an energy consumption model has not yet been studied, it is expected that \acp{SIM} will be more energy efficient compared to conventional digital transceivers.   

The study of \ac{SIM} is still in its infancy, and the main technological and modeling challenges have yet to be fully identified. For instance, the distance between the layers must be several wavelengths to validate the currently adopted cascade model. Power losses, frequency-selective effects, and signal distortion caused by the multiple layers and the bounding box require further investigation. A key limitation of \ac{SIM} is that only the final layer radiates, constraining the \ac{EM} wave characteristics to those achievable with a classical planar surface or array, while the role of the hidden layers is to perform an \ac{EM} transformation. 
Therefore, \acp{SIM} do not fully exploit the possibilities offered by \ac{EM} phenomena in antenna structures for processing and radiation, thereby inheriting some limitations of standard antenna arrays, as will be detailed later.

\subsection{Our Contribution}

In this paper, we investigate a 3D \ac{EM} antenna structure, termed \ac{DSA}, that offers significant flexibility by providing enhanced multifunctional processing capabilities directly at the \ac{EM} level. Specifically, a \ac{DSA} consists of a limited number of active antenna elements, each associated with an \ac{RF} chain, surrounded by a cloud of reconfigurable passive scatterers that interact with each other in the reactive near field. 
As anticipated, the \ac{DSA} generalizes concepts from reactively controlled arrays, their evolution into \acp{ESPAR}, and \acp{SIM}. Unlike \ac{SIM}, the DSA permits arbitrary distributions and couplings of passive scatterers, enabling simultaneous \ac{EM} processing and radiation. Furthermore, the same configuration of scatterers can perform different \ac{EM} processing functions, each associated with a distinct input to the \ac{DSA} \cite{Dar:C24}. It is worth noticing that \acp{ESPAR} have been mainly investigated considering only one input, narrowband signals, and limited structures \cite{KalKanPap:13,AleVlaCon:14,BucJuaKamSib:20}. 

We first characterize the input-output characteristic of the \ac{DSA} as a function of its configuration parameters and provide a fully analytical model using the impedance matrix of the structure.
With respect to \cite{Dar:C24}, we introduce a general, physically consistent framework for modeling a wideband \ac{DSA}, 
thereby enabling space-frequency processing capabilities. Notably, prior works on \acp{ESPAR} and \acp{SIM} have been limited to narrowband signals. 
In contrast to previous works, we relax the assumption of ideal reactive loads by introducing reconfigurable loads modeled using realistic varactor diodes, thereby capturing the effects of reduced antenna efficiency due to dissipated power and evaluating the corresponding $Q$-factor. 
We investigate the use of an ideal power matching network and propose a simplified, more practical alternative, demonstrating that the latter incurs no significant performance loss. This result effectively removes the need for a complex reconfigurable power matching network. It is worth noting that the models typically used to represent \ac{ESPAR} and \ac{SIM} do not consider the issue of the power matching network.
Building on the analytical expression of the frequency-dependent characteristics of the \ac{DSA} as a function of its configuration parameters, we formulate an optimization problem to design these parameters according to the desired multifunctional and multifrequency response. The optimization also considers an optional digital precoder, which might help the optimization of the \ac{DSA}.
The optimization strategy is then applied to 3 different use cases of interest: superdirective beamforming, multi-user \ac{MISO}, and \ac{MIMO} \ac{EM} precoding.
Numerical results for each use case demonstrate the great flexibility of the \ac{DSA} in performing various signal processing tasks. For instance, we will show that the tight coupling of \ac{DSA} elements allows for the realization of superdirectivity behavior independently of the beam's direction, in contrast to standard arrays where superdirectivity can only be achieved in the end-fire direction  \cite{IvrNos:10,HanYinMar:22}.   

\begin{figure*}[!t]
\centering 
\centering\includegraphics[width=2\columnwidth]{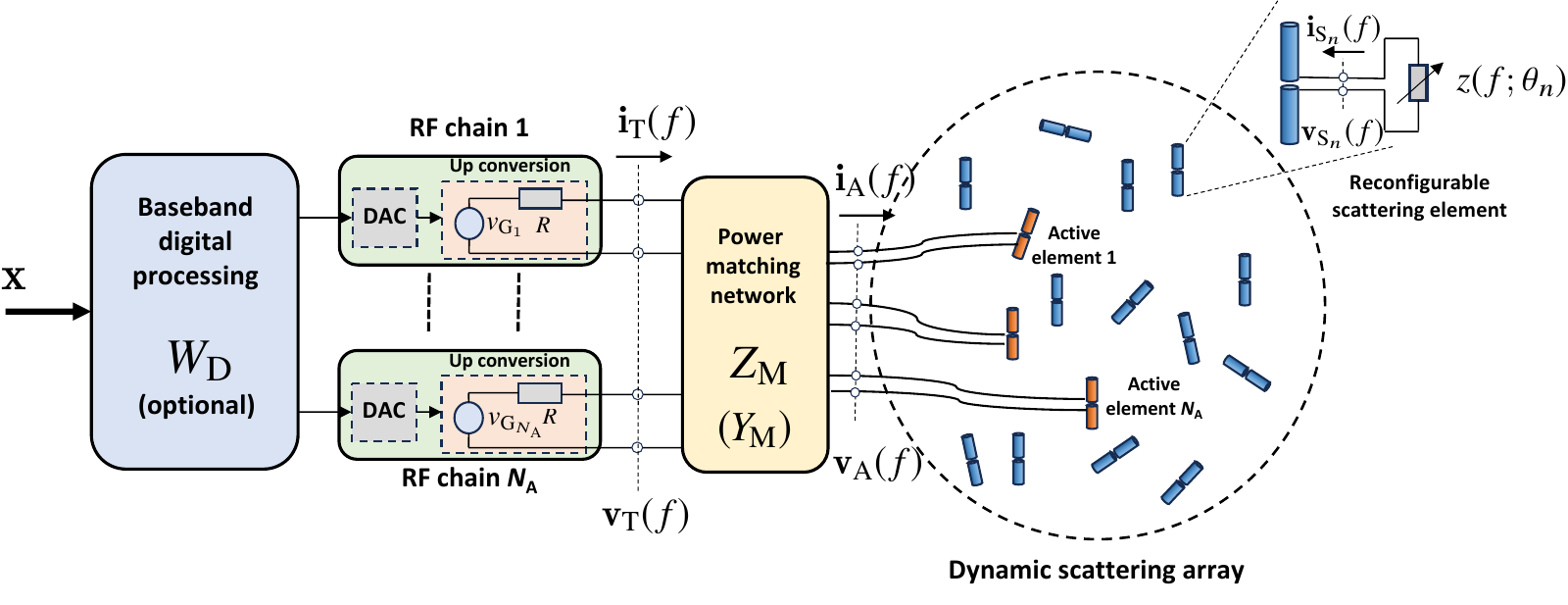}
\caption{Principle scheme of a dynamic scattering array with $\Na$ active antennas connected to a matching network, $\Na$ \ac{RF} chains, and an optional baseband digital processor.} 
\label{Fig:System}
\end{figure*}

We show that the state-of-the-art \ac{EM} processing structure, i.e., the \ac{SIM},  can be modeled as a particular case of a \ac{DSA}, characterized by a specific block lower bidiagonal impedance matrix structure. Compared with \ac{SIM}, in a \ac{DSA}, all the elements contribute jointly to the processing and radiation processes, thereby enhancing flexibility with more compact structures.
Indeed, in a \ac{SIM}, only the elements of the last layer radiate, and thus, the maximum achievable gain is proportional to their number, similar to a conventional antenna array. 
Furthermore, \acp{SIM} assume only forward propagation from one layer to the next, with no coupling between the elements of each layer. While this simplifies modeling and optimization significantly, it presents critical challenges in decoupling implementation and imposes size constraints. For example, in \cite{AnXuNgAleHuaYueHan:23} and \cite{HasAnDiRDebYue:24}, a separation of $5\lambda$ between each layer is considered, limiting the possibility of obtaining compact structures. In contrast, our numerical results demonstrate that with a \ac{DSA}, better performance can be achieved with a spacing between elements of $\lambda/4$, then with a significantly reduced size, thanks to the joint processing and radiation involving all the elements composing the \ac{DSA}.  
Specifically, numerical studies confirm that, for a given geometry, the \ac{DSA} outperforms a \ac{SIM} in approximating the desired \ac{EM} processing response while enabling the exploitation of superdirective capabilities.

\subsection{Notation and Definitions}
Boldface capital letters are matrices (e.g., $\boldA$), where $\boldI_N$ is the identity matrix of size $N$, $\boldzero_N$ is the zero matrix of size $N$. With reference to a generic matrix $\boldA$,  $a_{n,m}=[\boldA]_{n,m}$ represents its $(n,m)$th element, whereas  $\boldA\transpose$ and $\boldA\ctranspose$ indicate the transpose and the conjugate transpose of $\boldA$, respectively. We denote with $\boldA^{\dag}$ the Moore-Penrose inverse (pseudo-inverse) and with $\left \|  \boldA  \right \|_{\text{F}} $ the Frobenius norm of matrix $\boldA$. We indicate with $\Real{z}$, $\Imag{z}$, and $z^*$, respectively,  the real part,  imaginary part, and complex conjugate of the complex number $z$, and $\jmath$ the imaginary unit.
The statistical expectation of a random variable $x$ is indicated as $\EX{x}$.
Denote with $\eta=377\,$Ohm the free-space impedance and $c$  the speed of light.

\subsection{Paper Organization}
The rest of the paper is organized as follows: 
Sec.~\ref{Sec:Model} introduces the general framework to model the  \ac{DSA}.
The optimization of the parameters characterizing the \ac{DSA} is addressed in general in Sec.~\ref{Sec:Optimization}, and further declined in 3 use cases. Numerical results associated with the 3 use cases are provided in Sec.~\ref{Sec:NumericalResults}. Finally, the conclusions are drawn in Sec.~\ref{Sec:Conclusion}.

\section{\acf{DSA} Modeling}
\label{Sec:Model}

With reference to Fig.~\ref{Fig:System}, a \ac{DSA} is composed of $\Na$ active antenna elements, each of them connected to an \ac{RF} chain, surrounded by a cloud of $\Ns$ reconfigurable passive scatterers interacting in the reactive near field for a total of $N=\Na+\Ns$ elements. In the figure, elements are seemingly deployed at random to emphasize that the proposed framework is applicable to arbitrary 2D and 3D geometries.
The frequency-dependent response of the \ac{DSA}  can be changed by loading each antenna element using a reconfigurable load of frequency-dependent impedance $z(f; \theta)$, which is a function of parameter $\theta$  that affects the impedance according to the specific load implementation technology, as it will be detailed in Sec. \ref{Sec:Varactor}. 
Denote with  $\btheta=\left [\theta_1, \theta_2, \ldots, \theta_{\Ns} \right ]$ the set of reconfigurable parameters of the $\Ns$ loads and with $\bZs=\diag{z(f;\theta_1), z(f;\theta_2), \ldots , z(f;\theta_{\Ns})} $ the diagonal matrix collecting the corresponding set of loads' impedances.  
 
The \ac{DSA} can be modeled as a linear frequency-selective $N$-port network. 
In a compact description, we define  $\bva(f) \in \mathbb{C}^{\Na \times 1}$ and $\bia(f) \in \mathbb{C}^{\Na \times 1}$ the complex voltage and current envelopes  (in the following denoted simply as voltages and currents) at the ports of the $\Na$ active antennas at the generic symbol time,\footnote{For notation convenience, we omit the time dependence. } 
and $\bvs(f) \in \mathbb{C}^{\Ns \times 1}$ and $\bis(f) \in \mathbb{C}^{\Ns \times 1}$ the voltages and currents at the ports of the $\Ns$ scatterers. Since voltages and currents carry information, they must be treated as random variables for a given frequency.
We collect all the currents and voltages at the $N$ ports of the \ac{DSA} in the vectors 
  \begin{align} 
&\bi(f)=\left [
\begin{array}{c}
 \bia(f)      \\
 \bis (f)         
\end{array}
   \right ]  \in \mathbb{C}^{N \times 1}
 &\bv(f)=\left [
\begin{array}{c}
 \bva(f)      \\
 \bvs(f)          
\end{array}
   \right ] \in \mathbb{C}^{N \times 1} \, .
\end{align}

All the interactions between the elements of the \ac{DSA} are captured by the impedance matrix $\bZ  \in \mathbb{C}^{N \times N}$, which does not depend on the reconfigurable loads, and relates the voltages and currents of the $N$ ports as $\bv(f) =\bZ \, \bi(f)$. It can be computed once during the design of the \ac{DSA} according to the technology adopted. In Sec. \ref{Sec:NumericalResults}, an example will be given. 
In particular, the $(m,n)$th element of $\bZ$ represents the mutual coupling coefficient between the $n$th and $m$th elements obtained as the ratio between the open-circuit voltage observed at the $m$th port and the excitation sinusoidal current at frequency $f$ applied to the $n$th port supposing that the remaining ports are kept open (no current flow) \cite{BalB:16}.   It is convenient to divide the impedance matrix  $\bZ$ into the sub-matrices $\bZaa$, $\bZas$, $\bZsa$, and $\bZss$ so that we can write
\begin{align} \label{eq:vavs}
\left [
\begin{array}{c}
 \bva(f)      \\
 \bvs(f)          
\end{array}
   \right ]=\left [
\begin{array}{cc}
 \bZaa & \bZas     \\
 \bZsa & \bZss          
\end{array}
   \right ] \cdot \left [
\begin{array}{c}
 \bia(f)      \\
 \bis(f)          
\end{array}
   \right ] \, .
\end{align}

The voltages $\bva(f)$ and currents $\bia(f)$ at the input ports of the \ac{DSA} are related by the input impedance of the \ac{DSA} $\bZa \in \mathbb{C}^{\Na \times \Na}$ as $\bva(f)=\bZa \, \bia(f)$. 
By elaborating \eqref{eq:vavs} and considering that at the scatterers' ports it is $\bvs(f)=-\bZs \, \bis(f)$, we obtain
\begin{align} \label{eq:bZa}
\bZa=\bZaa-\bZas  \left (\bZss+\bZs  \right )^{-1}  \bZsa 
\end{align}
which depends on parameters $\btheta$ through the scatterers' loads $\bZs$.
For further convenience, possible dissipative components in the antenna elements are accounted for as additional resistive terms in the loads $\bZs$. As a consequence, the radiated power $\Prad(f)$ at frequency $f$ can be evaluated as
\begin{align}
	\Prad(f)=\EX{\bi(f)\ctranspose \, \Real{\bZ} \, \bi(f)} 
\end{align}
whereas the reactive power is 
\begin{align}
	\Preact(f)=\EX{\bi(f)\ctranspose \, \Imag{\bZ} \, \bi(f)} 
\end{align}
from which the $Q$-factor of the structure can be obtained as $Q(f)=\Preact(f)/\Prad(f)$. The inverse of the $Q$-factor can be used as a rough estimate of the bandwidth of the \ac{DSA} when $Q\gg 1$\cite{BalB:16}. 

The $\Na$ active elements are connected to the $\Na$ \ac{RF} chains, including \ac{DAC} and up-conversion stages, through a $2\Na$-port power matching network whose purpose is to maximize the power transfer between the \ac{RF} chains and the \ac{DSA}. 
The available transmitted power at the output of the \ac{RF} chains is 
\begin{align}
	\Ptx(f)=\frac{\EX{\bvg(f)\ctranspose \, \bvg(f)}}{4 R} 
\end{align}
with $\bvg(f)=\left [\vg_1(f), \vg_2(f), \ldots , \vg_{\Na}(f)\right ]\transpose$ representing the open-circuit voltages of the RF chains and $R$ being the internal resistance of each chain. It coincides with the power transferred to the \ac{DSA} only when a perfect power-matching network is introduced. 

In the following, we consider two different implementations of the network connecting the \ac{DSA} and the $\Na$ \ac{RF} chains: a) Perfect matching network; b) Simplified (not perfect) matching network.

\subsection{Perfect Matching Network}

In this implementation, the relationship between the voltages and the currents present at the ports of the matching network is given by  
\begin{align} \label{eq:vtva}
\left [
\begin{array}{c}
 \bvt(f)      \\
 \bva(f)          
\end{array}
   \right ]=\bZm{f} \, \left [
\begin{array}{c}
 \bit(f)      \\
 -\bia(f)          
\end{array}
   \right ]
\end{align}
where, assuming no losses in the power matching network, $\bZm{f}$ is equal to  \cite{IvrNos:10}
\begin{align} \label{eq:bZm}
\bZm{f}=\left [
\begin{array}{cc}
 \boldzero_{\Na} & -\jmath \sqrt{R} \, \Real{\bZa}^{\frac{1}{2}}      \\
 -\jmath \sqrt{R} \, \Real{\bZa}^{\frac{1}{2}} & -\jmath \Imag{\bZa}           
\end{array}
   \right ] 
\end{align}

As it can be noticed from \eqref{eq:bZm}, the matching network depends on the \ac{DSA} input impedance $\bZa$ and hence on the parameters $\btheta$. Therefore, for a given configuration of the \ac{DSA}, the matching network should be changed accordingly. This is possible using, for instance, adaptive matching networks \cite{AliShu:20}. In case a fixed network is employed instead, then not all the available power would be transferred to the \ac{DSA} with a consequent reduction of efficiency. An alternative and more practical solution is proposed in Sec. \ref{Sec:SimpleMatching}.
With the perfect matching network, it turns out that  $\bvt(f)=R\, \bit(f)$ and 
the actual power transferred to the \ac{DSA} coincides with the available power
\begin{align}
	\Pa(f)=\frac{\EX{\bvt(f)\ctranspose \, \bvt(f)}}{R}=\Ptx(f) \, .
\end{align}

From \eqref{eq:vtva} and \eqref{eq:bZm} it is
\begin{align} \label{eq:bia}
\bia(f)=\frac{1}{\jmath 2 \sqrt{R}} \, \Real{\bZa}^{-\frac{1}{2}}   \bvg(f)  \, .
\end{align}
Moreover, we have
\begin{align} \label{eq:bi}
\bi(f)
=\left [
\begin{array}{c}
 \boldI_{\Na}    \\
 -\left ( \bZss + \bZs \right )^{-1} \bZsa        
\end{array}
   \right ] \cdot 
  \bia(f) 
  \, .
\end{align}

By combining \eqref{eq:bia} and \eqref{eq:bi}, we obtain a compact relationship between the open-circuit voltages of the RF chains $\bvg (f)$ and the total current $\bi(f)$ flowing in the \ac{DSA}
\begin{equation} \label{eq:ik}
\bi(f)
=\bWem{f} \, \bvg (f) 
\end{equation}
where 
\begin{align} \label{eq:Wa}
& \bWem{f}
= \nonumber \\
& \frac{1}{\jmath 2 \sqrt{R}} \, \left [
\begin{array}{c}
  \Real{\bZa}^{-\frac{1}{2}}    \\
 - \left ( \bZss + \bZs \right )^{-1} \bZsa   \, \Real{\bZa}^{-\frac{1}{2}}      
\end{array}
   \right ]  
\end{align}
accounts for the \ac{EM}-level signal processing operated by the reconfigurable \ac{DSA} as a function of the parameters $\bpsi=\{\btheta\}$.

The the radiated power is $\Prad(f)=\etad(f) \,  \Ptx(f)$, where $\etad(f)=\Prad(f)/\Pa(f)=1-\Pd(f)/\Pa(f)$ represents the power loss due to dissipation at frequency $f$ and $\Pd(f)=\EX{\bis(f) \ctranspose \, \Real{\bZs} \bis(f) }$ is the dissipated power. 

\begin{figure}[!t]
\centering 
\centering\includegraphics[width=0.4\columnwidth]{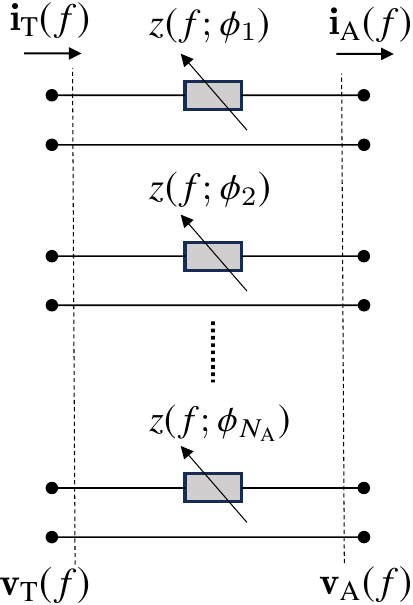}
\caption{Simplified matching network.} 
\label{Fig:NoMatching}
\end{figure}

\subsection{Simplified Matching Network}
\label{Sec:SimpleMatching}

We propose the simplified matching network represented in the scheme shown in Fig. \ref{Fig:NoMatching}, 
where $\Na$ additional reconfigurable impedances $\bZl=\left [z(f; \phi_1), z(f; \phi_2), \ldots, z(f; \phi_{\Na}) \right ]^{\transpose}$ are added at the input of the active elements of the \ac{DSA} so that the matching task enters the optimization process. The vector $\boldsymbol{\phi}=[\phi_1, \phi_2, \ldots, \phi_{\Na}  ]$ collects the corresponding parameters to be optimized. 
The relationship between the voltages and the currents present at the ports of the matching network can be expressed through the admittance matrix\footnote{In this case, the corresponding impedance matrix cannot be defined.} given by  
\begin{align} \label{eq:vtvaY}
\left [ \begin{array}{c}
 \bit(f)      \\
 -\bia(f)          
\end{array}
   \right ]
=\bYm{f} \,
\left [
\begin{array}{c}
 \bvt(f)      \\
 \bva(f)          
\end{array}
   \right ]
\end{align}
where
 \begin{align} \label{eq:bYm}
\bYm{f}=\bZl^{-1} \left [
\begin{array}{cc}
\boldI_{\Na} & -\boldI_{\Na}      \\
 -\boldI_{\Na} & \boldI_{\Na}     
\end{array}
   \right ]  \, . 
\end{align}
Using \eqref{eq:vtvaY} and \eqref{eq:bYm}, the open-circuit voltage of the \ac{RF} chains is $\bvg(f)= \boldQ(f,\bphi)\, \bia$, where $\boldQ(f,\bphi)=\bZa+\bZl+ R \, \boldI_{\Na}$. The relationship between $\bvg (f)$  and $\bi(f)$ is still given by \eqref{eq:ik}, where now 
\begin{align} \label{eq:Wano}
& \bWem{f} = \left [
\begin{array}{c}
  \boldQ(f, \bphi)^{-1}    \\
 - \left ( \bZss + \bZs \right )^{-1} \bZsa   \, \boldQ(f,\bphi)^{-1}  
\end{array}
   \right ]  
\end{align}
accounts for the \ac{EM}-level signal processing operated by the reconfigurable \ac{DSA} as a function of the overall set of parameters $\bpsi=\{\btheta; \bphi\}$.

In this case,  the actual power transferred to the \ac{DSA} is $\Pa(f)=\etam(f) \, \Ptx(f)$, where $\etam(f)$ is the impedance mismatch loss ($\etam(f)=1$ with the perfect matching network), and the radiated power is $\Prad(f)=\etad(f) \, \etam(f) \, \Ptx(f)$, where  $\etad(f)=1-\Pd(f)/\Pa(f)$, with $\Pd(f)=\EX{\bis(f) \ctranspose \, \Real{\bZs} \bis(f) } +\EX{\bia(f) \ctranspose \, \Real{\bZl} \bia(f) } $ accounting for the power dissipated in the \ac{DSA} and the matching network.

\subsection{Digital Processing}
We also foresee an optional baseband linear processing block implemented in the digital domain (digital precoder), described by the matrix $\bWdo \in \mathbb{C}^{\Na \times \Na}$, whose input is represented by the information vector $\boldx(f) \in \mathcal{C}^{\Na \times 1}$ to be transmitted. The matrix $\bWdo$  relates the transmitted information vector $\boldx(f)$ and the open-circuit voltages $\bvg(f) $ of the \ac{RF} chains, i.e., $\bvg(f) = \bWdo\, \boldx(f) $.  In the following, we impose the arbitrary normalization $\| \bWdo \|_{\text{F}}=\sqrt{4 R  \Na}$. Such a normalization ensures that $\EX{\boldx(f) \ctranspose \, \boldx(f) }=\Ptx(f) $, where $\Ptx(f)$ is the (available) transmitted power, as typically assumed in the signal processing community.

The reconfigurable parameters $\bpsi$ along with the elements of matrix $\bWdo$ represent the set of parameters to be optimized to obtain the desired processing functionality, as will be described later.

\subsection{Particular Case: SIM Modeling}

In this section, we show how the general framework illustrated in the previous section can be made particular to model a \ac{SIM}.

Suppose the $\Ns$ reconfigurable scattering elements are organized in layers, where the $l$th layer is composed of $N_l$ elements, such that $\sum_{l=1}^L N_l=\Ns$. A typical assumption for modeling a \ac{SIM} is to consider no coupling between elements belonging to the same layer, and only forward propagation, meaning that coupling takes place only between each element of layer $l$ and the elements of layer $l+1$, but not vice versa. This translates into the following block lower bidiagonal structure of the impedance matrix of the \ac{DSA}
\begin{align} \label{eq:ZSIM}
\bZ= \left [
\begin{array}{ccccc}
  \bZaa & 0 & 0 & \ldots & 0    \\
  \boldB_1(f) & \boldA_1(f) & \boldzero & \ldots & \boldzero \\
  \boldzero & \boldB_2(f) & \boldA_2(f) & \ldots & \boldzero\\
  \vdots &\vdots & \vdots & \ddots & \vdots\\
  \boldzero & \boldzero & \boldzero & \ldots & \boldA_L(f)\\
\end{array}
   \right ]  
\end{align} 
where $\boldA_l(f) \in \mathbb{C}^{N_l \times N_l}$, $l=1,2, \ldots , L$, is a diagonal matrix representing the self-impedance of the scattering elements belonging to layer $l$. Also $\bZaa \in \mathbb{C}^{\Na \times \Na}$ is, in this case, a diagonal matrix associated to the self-impedance of the $\Na$ active antennas. 
Matrix   $\boldB_l(f) \in \mathbb{C}^{N_l \times N_{l-1}}$, for $l=1,2, \ldots , L$, accounts for the (forward) coupling between layer $l$ and layer $l-1$ (when $l=1$, the coupling is with the active antennas, and $N_{l-1}=\Na$).
From simple matrix manipulations, since $\bZas=\boldzero$, the \ac{DSA} input matrix is simply given by $\bZa=\bZaa$, which does not depend on the \ac{DSA} configuration $\btheta$. 
Therefore, there is no need for an adaptive power matching network (we set $\bZl=0$). For instance, the $\Na$ active elements can be matched to the \ac{RF} chain internal impedance $R$, thus leading to $\bZa=R \, \boldI_{\Na}$ and $\boldQ(f,\bphi)=\boldQ=2\,R \, \boldI_{\Na}$.
Denote with $\bi_l(f) \in \mathbb{C}^{N_l \times 1} $, for $l=1,2, \ldots , L$, the currents flowing in the elements of the $l$th layer. It follows that

\begin{align} \label{eq:W1}
& \left [ \begin{array}{c}
 \bia(f)      \\
 \hline
 \bi_1(f) \\
 \bi_2(f) \\   
 \vdots \\
 \bi_L(f)       
\end{array}
   \right ] = \left [
\begin{array}{c}
  \boldQ^{-1}    \\
  \hline
 - \left ( \bZss + \bZs \right )^{-1} \bZsa   \, \boldQ^{-1}  
\end{array}
   \right ]  \,  \bvg(f)     \, .
\end{align}

In a typical \ac{SIM}, only the last layer is designed to radiate outside the structure; therefore in this case, the matrix $\bWem{f} \in \mathbb{C}^{N_L \times \Na}$ is the submatrix in \eqref{eq:W1} which puts in relationship the input voltages $\bvg(f)$ with only the currents $\bi_L(f)$ of the last layer. By exploiting the particular block lower bidiagonal structure of $ \bZss + \bZs$, and by applying the iterative Schur complement, we obtain
\begin{align} \label{eq:WaSIM}
 \bWem{f} =&
 (-1)^L  \boldB_L(f) \, \bPhi_L(f;\btheta) \ldots \nonumber \\
 & \,  \ldots   \bPhi_2(f;\btheta)  \, \boldB_2(f) \, \bPhi_1(f;\btheta) \, \boldB_1(f) \, \boldQ^{-1}  
\end{align}
having defined the diagonal matrix
\begin{align} 
\bPhi_l(f;\btheta)=\left ( \boldA_l(f)+ \diag{z(f;\theta_{p_l+1}) \, , \ldots , z(f;\theta_{p_{l+1}}) } \right )^{-1}
\end{align}
for $l=1,2, \ldots, L$, with $p_l=\sum_{i=1}^{l-1} N_i$.
Equation \eqref{eq:WaSIM} reveals that the typical modelling of a \ac{SIM} can be obtained as a particular case of our general framework. The characteristic chain structure of \eqref{eq:WaSIM} enables the adoption of gradient descent algorithms based on error backpropagation, similar to those used in deep neural networks \cite{AnYueGuaDiRDebPooHan:24}.  
Nevertheless, while the characteristics of the  \ac{SIM} allow for simplified optimization schemes, as already pointed out, only the last layer contributes to the radiation, thus limiting the flexibility of the \ac{SIM} compared to a \ac{DSA}. 
It is worth underlining that model \eqref{eq:WaSIM} is based on relatively oversimplified assumptions that do not take into account nonidealities such as couplings between elements of the same layer or backscattering from subsequent layers. Such non-idealities can be easily included in our general framework through the impedance matrix  \eqref{eq:ZSIM}, which, however, would lose its block lower bidiagonal structure, and a more general optimization scheme would be required, as proposed in the next section.

\section{DSA Optimization}
\label{Sec:Optimization}

Consider now $T$ test points located in positions $\boldt_t$, with $t=1,2, \ldots , T$, and that in each test point, a conventional receiving wideband antenna is used to receive the signal with gain $\Gr$. As commonly done in the literature, we assume that the test points are located in the radiative region of the \ac{DSA} and the receiving antennas do not affect the \ac{DSA}. 
As examples, the set of $T$ antennas could represent a conventional receiving antenna array of a \ac{MIMO} system or users located in different positions.  
In addition, we suppose a multicarrier signaling with total bandwidth $W$ using $K$ subcarriers at frequencies $f_k=f_0+(k-1) W/K$, for $k=1,2, \ldots K$. To be general, we consider that a different digital processing matrix $\bWdk$, for $k=1,2, \ldots, K$, is applied to each subcarrier. 

The useful component (i.e., without noise) of the received signal at the test positions at frequency $f_k$ is 
\begin{align} \label{eq:by}
	\boldy(f_k)& =\bHc{f_k} \, \bi(f_k)+\boldn(f_k)\nonumber \\
	&=\bHc{f_k} \, \bWem{f_k} \, \bWdk \, \boldx(f_k)  +\boldn(f_k) \nonumber \\
	 &=\boldH_k(\bpsi, \bWdk) \, \boldx(f_k) +\boldn(f_k)
\end{align}
where $\boldn(k)$ is the zero-mean complex Gaussian \ac{AWGN} with convariance matrix $\sigma^2 \boldI_K$,  $\bHc{f_k}  \in \mathbb{C}^{T \times N}$ is the transimpedance matrix of the radio channel accounting for the propagation effects, and $\boldH_k(\bpsi, \bWdk) \in \mathbb{C}^{T \times \Na}$ represents the end-to-end baseband equivalent channel matrix as commonly defined in signal processing.

In this section, we illustrate a possible optimization strategy of the \ac{DSA} along with its application to 3 different use cases. Numerical results associated with each use case will be reported in Sec.~\ref{Sec:NumericalResults}.

\subsection{Joint Optimization of the DSA and the Digital Precoder}

For a given propagation scenario characterized by the transimpedance matrix $\bHc{f} $, $T$ test points, and $K$ frequencies $\{f_k\}$, suppose we fix a specification $\bHoptk \in \mathbb{C}^{T \times \Na}$ on the desired end-to-end channel matrix (i.e., our desired processing). The reconfigurable  parameters of the \ac{DSA} and the digital precoder can be obtained by solving the following constrained optimization problem:
\begin{align}
\label{eq:opt}
& 
\minimize{\bpsi, \, \balpha , \, \left \{ \bWdd{k}  \right \}  } \sum_{k=1}^K \left \|  \alpha_k \, \bHc{f_k} \, \bWem{f_k} \, \bWdk - \bHoptk  \right \|_{\text{F}}^2 \\
& s.t. \quad \| \bWdk \|_{\text{F}}^2=4 R  \Na  \, \, \quad k=1,2, \ldots , K	\nonumber
\end{align}
where $\balpha=\{ \alpha_1, \alpha_2, \ldots, \alpha_K\}$ with the scalars $\alpha_k \in \mathbb{R}$ accounting for the possible lack in the link budget to achieve $\bHoptk$ that has to be eventually coped with increased transmitted power. It is worth noting that the $n$th column of $\bHoptk$ represents the desired response of the \ac{DSA} at the $T$ test points to the $n$th input data stream $x_n(f_k)$ at frequency $f_k$ (alias the $n$th processing function). Therefore, the minimization of \eqref{eq:opt} leads to a configuration for which the \ac{DSA} approximates the multi-functional processing of the inputs at the \ac{EM} level.    

Unfortunately, in general, the constrained optimization problem in \eqref{eq:opt} is not convex thus making its numerical solution more challenging. A possible approach to translating it into an unconstrained optimization problem is to resort to the following alternate optimization procedure:

\begin{itemize}
\item {STEP 0:}
We set the candidate vector $\hat{\bpsi}$  to an initial guess value, for instance randomly chosen.

\item {STEP 1:}

For a fixed $\hat{\bpsi}$, it is possible to find in closed form the values of $\bWdk$ and $\alpha_k$ that minimize the objective transfer function and satisfy the constraint in \eqref{eq:opt}. In particular, we solve the following equation
\begin{align}
\alpha_k \, \bHc{f_k} \, \bWemhat{f_k} \, \bWdk = \bHoptk
\end{align}
of which the minimum norm solution is
\begin{align}
 \alpha_k \bWdk  =  \left (\bHc{f_k}\,  \bWemhat{f_k}    \right )^{\dag} \bHoptk \, .
\end{align}

The corresponding values of $\hat{\alpha}_k$ and $\bWdhatk$ are, therefore,
\begin{align} \label{eq:hatalpha}
 \hat{\alpha}_k   = \frac{1}{\sqrt{4 R \Na} } \left \| \left (\bHc{f_k}\,  \bWemhat{f_k}    \right )^{\dag} \bHoptk \right \|_{\text{F}}
\end{align}
and
\begin{align}
  \bWdhatk  = \frac{1}{\hat{\alpha}_k } \left (\bHc{f_k} \,  \bWemhat{f_k}    \right )^{\dag} \bHoptk \, .
\end{align}

\item {STEP 2:} 

Starting from the candidates  $\hat{\alpha}$ and $\bWdhatk$ from STEP 1, the following unconstrained minimization problem involving only $\bpsi$ is solved numerically
\begin{align}
\label{eq:opt1}
& \hat{\bpsi} =\argmin{\bpsi}  \sum_{k=1}^K \left \|  \hat{\alpha}_k \, \bHc{f_k} \, \bWem{f_k} \, \bWdhatk - \bHoptk  \right \|_{\text{F}}^2  \, .
\end{align}


\item STEP 3: Repeat STEPS 1 and 2 for a given number $N_{\text{alt}}$ iterations or until no significant variations of the parameters take place.
\end{itemize} 

\begin{figure}[!t]
\centering 
\centering\includegraphics[width=1\columnwidth]{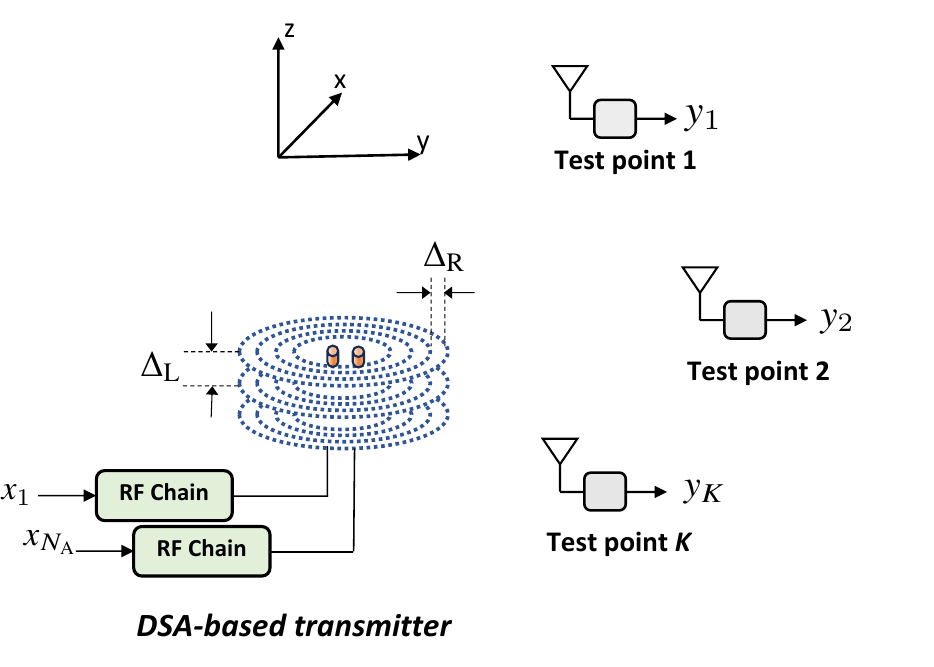}
\caption{Use case 1: Single RF chain DSA for superdirective beamforming.} 
\label{Fig:Example1}
\end{figure}

\subsection{Use Case 1: Superdirective Beam Forming}
\label{Sec:Example1}

The inclusion of the digital precoder, while not strictly necessary, may aid in the optimization process of the \ac{DSA} by handling a portion of the processing when $\Na>1$. If omitted, we set $\bWdhatk=\bWdk= \sqrt{4  R} \boldI_{\Na}$, and only proceed with STEP 2 once.
Of course, alternative optimization problems apart from \eqref{eq:opt} can also be considered, such as those involving the $Q$-factor as an optimization target or constraint.
In many applications (e.g., beam steering, as in use case 1 below), the optimization process can be performed offline, computed once to create a dictionary of various \ac{DSA} parameters for different steering directions of interest or, more broadly, different beam forms. In such cases, the complexity of the optimization algorithm is not a concern.
However, in other scenarios, performing optimization offline may not be feasible, necessitating online optimization based, for example, on the \ac{CSI} of $\bHc{f_k}$. In this situation, developing efficient algorithms to minimize \eqref{eq:opt1}, potentially considering the structure of $\bWemhat{f_k}$, becomes pertinent. This aspect exceeds the scope of this paper and is reserved for future investigations.
In the numerical results, the standard numerical tool based on the quasi-Newton method \cite{FleB:80} with $N_{\text{i}}$ iterations is utilized to minimize \eqref{eq:opt1}.

\begin{figure}[!t]
\centering 
\centering\includegraphics[width=1\columnwidth]{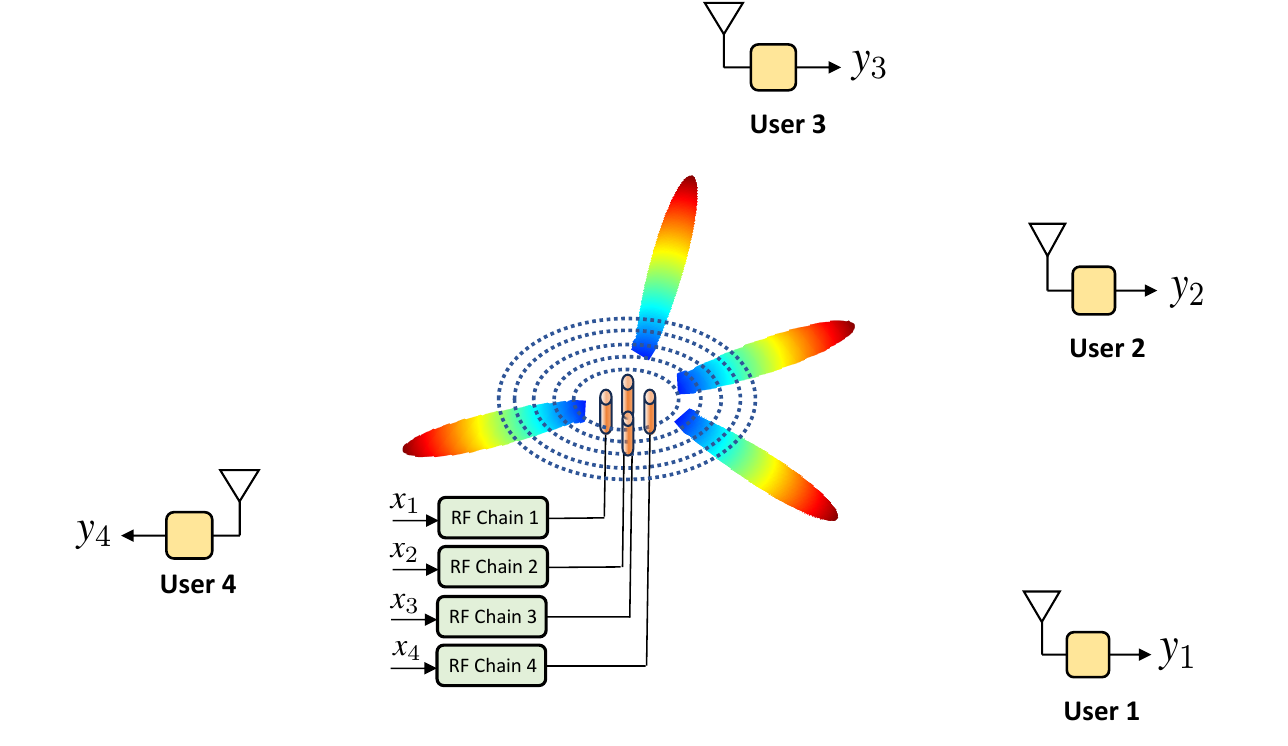}
\caption{Use case 2: Multi-user downlink MISO with a DSA.} 
\label{Fig:Example2}
\end{figure}

Suppose we want to design a single-input \ac{DSA} ($\Na=1$, no digital precoding) realizing a generic radiation diagram associated with subvarrier $k$, for instance, imposing that at the test locations $\boldt_t$,  the channel matrix $\left [\bHoptk \right ]_{t,1}$, $t=1,2, \ldots, T$, assumes some desired values (see Fig.~\ref{Fig:Example1}). 
For instance, one might define a uniform set of test points $\boldt_t=[d\sin \phi_t ,0 ,d \cos \phi_t]\transpose$ on the $x-y$ plane a distance $d$ in the far-field region of the \ac{DSA}, with $\phi_t=2 \pi t / T$, and maximize the radiation diagram at the $t=\tilde{t}$th direction $\phi_{\tilde{t}}$ (beam steering). This can be achieved by setting   $\left [\bHoptk \right ]_{t,1}=\sqrt{\Prx}$, for $t=\tilde{t}$, and zero otherwise, being $\Prx$ the desired received power.  
As it will be shown in the numerical results, a dedicated radiation diagram can be obtained for each subcarrier. It is worth noticing that, in principle, only the optimization at the single test point $\boldt_{\tilde{k}}$, i.e., $T=1$, would suffice to obtain the desired beam steering result. However, the addition of many other test points set to zero helps the minimization algorithm to speed up and avoid the convergence to bad local minima.

\subsection{Use Case 2: Multi-user \ac{MISO}} 
\label{Sec:Example2}

In this second use case, we consider a multi-user \ac{MISO} downlink scenario shown in Fig.~\ref{Fig:Example2},  where an  \ac{DSA} with  $\Na$ \ac{RF} chains (one per user) has to be designed to serve simultaneously $T=\Na$ single-antenna users located in generic positions $\boldt_t$, $t=1,2, \ldots, T$, by maximizing the \ac{SINR} at each user according to the zero-forcing criterium \cite{TseVis:B05}. Users can be both in the radiating near-field and far-field regions of the \ac{DSA}.

Given the channel transimpedance matrix $\bHc{f_k}=\left [[\bhc{f_k}]_1 ; [\bhc{f_k}]_2 ; \ldots ; [\bhc{f_k}]_{\Na}  \right ]$, where $[\bhc{f_k}]_t$ is the $N \times 1$ channel  vector associated with the $t$th user, the zero-forcing precoding matrix is $\boldV_k=\beta_k \, \bHc{f_k}^{\dag}$, where $\beta_k$ is a normalization factor to make $\boldV_k$ unitary, i.e., $ \boldV\ctranspose_k \boldV_k =\boldI_N$ \cite{TseVis:B05}. Therefore, the \ac{DSA} and the digital precoder must be designed such that
 \begin{equation} \label{eq:ZF}
 	 \bWem{f_k} \, \bWdk=\boldV_k=\beta_k \, \bHc{f_k}^{\dag} 
 \end{equation}
 leading to
 \begin{align}  
\boldy(f_k)&=\bHc{f_k} \,  \bWem{f_k} \, \bWdk \, \boldx(f_k) +\boldn(f_k) \nonumber \\
&=\bHc{f_k} \, \bHc{f_k}^{\dag} \, \boldx(f_k)+\boldn(f_k) \, .
\end{align}

This is equivalent to setting $\bHoptk=\bHc{f_k} \, \bHc{f_k}^{\dag}$ in the optimization problem in \eqref{eq:opt}.

 \begin{figure}[!t]
\centering 
\centering\includegraphics[width=1\columnwidth]{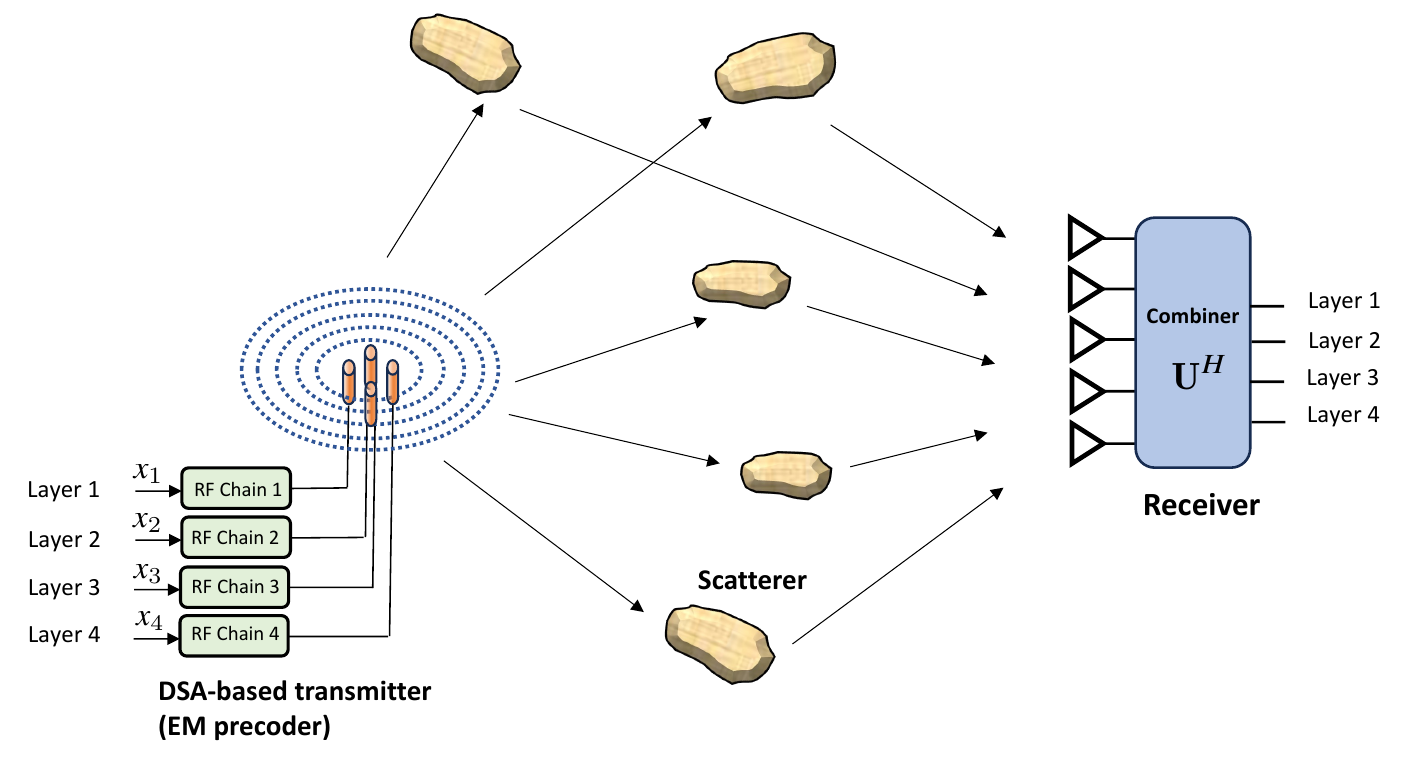}
\caption{Use case 3: DSA working as a MIMO EM precoder.} 
\label{Fig:Example3}
\end{figure}

\subsection{Use Case 3:  \ac{MIMO} EM Precoder } 
\label{Sec:Example3}

In this use case, we consider a \ac{MIMO} communication link with a receiving conventional \ac{ULA} with $T$ elements spaced at $\lambda/2$ illustrated in Fig.~\ref{Fig:Example3}. The purpose is to design the \ac{DSA} in such a way an optimal multi-layer \ac{MIMO} link is established between the transmitter and the receiver. It is well known from \ac{MIMO} theory that for a given generic propagation scenario with channel transimpedance matrix $\bHc{f_k}$ characterized by rank $r$, up to $r$ parallel orthogonal links (layers) can be established between the transmitter and the receiver on which $r$ independent data stream can be transmitted \cite{TseVis:B05}. To exploit all of them, it must be $\Na=r$ and the \ac{DSA} must implement a suitable precoding strategy, i.e., act as an \emph{\ac{EM} precoder}.  
  
Let us introduce the \ac{SVD} of the channel transimpedance
 \begin{equation}
 	\bHc{f_k}=\boldU_k\, \bLambda_k\,  \boldV_k\ctranspose
 \end{equation}
 where $\boldU_k$ and $\boldV_k$ contain the left and right eigenvectors, respectively, and $\bLambda_k$ is a diagonal matrix gathering the singular values of the channel transimpedance matrix $\bHc{f_k}$.
 A  typical \ac{MIMO} setup, assuming the \ac{CSI} is available at the transmitter, requires a precoding operation $\boldV_k$ at the transmitter and a combining operation $\boldU\ctranspose_k$ at the receiver such that the channel is diagonalized, i.e., the end-to-end channel matrix is proportional to the diagonal matrix $\bLambda_k$. Assuming the receiver performs the combining operation $\tilde{\boldy}(f_k)=\boldU\ctranspose_k\, \boldy(f_k)$, the \ac{MIMO} channel is diagonalized if we design the \ac{DSA} implementing the precoding $\boldV_k$ by setting 
  \begin{equation} \label{eq:SVD}
 	 \bWem{f_k} \, \bWdk=\boldV_k
 \end{equation}
 or, equivalently,
 \begin{equation}
 	\bHoptk=\boldU_k \, \bLambda_k \, .
 \end{equation}
 In this manner, it results
 \begin{align}
 	\tilde{\boldy}(f_k)=& \boldU\ctranspose_k \, \boldH \left (\bpsi, \bWdk \right ) \, \boldx(f_k) +\boldn(f_k) \nonumber \\
	&=\boldU\ctranspose_k \, \bHc{f_k} \,  \bWem{f_k} \, \bWdk \, \boldx(f_k) +\boldn(f_k)\nonumber \\
	&= \bLambda_k \,  \boldV\ctranspose_k \,  \boldV_k \, \boldx(f_k)+\boldn(f_k)=\bLambda_k \, \boldx(f_k)+\boldn(f_k)
 \end{align}
and $\Na$ parallel communication layers can be established. The designed \ac{DSA} implements the optimal precoding at the \ac{EM} level, thus saving $2 N^2$ multiplications and sums in the digital domain, by using no more than $\Na=r$ RF chains (i.e., the minimum possible value), which is much simpler and less energy consuming than any conventional full digital or hybrid solution. 

In our numerical investigations, we discovered that a more robust optimization is obtained by solving the following equivalent problem instead of \eqref{eq:opt1}
\begin{align}
\label{eq:opt2}
\minimize{\bpsi, \, \balpha , \, \left \{ \bWdd{k} \right \}    }
 \sum_{k=1}^K \left \|  \hat{\alpha}_k \, \boldU\ctranspose_k \bHc{f_k} \, \bWem{f_k} \, \bWdk - \bLambda_k  \right \|_{\text{F}}^2  
\end{align}
 that consists in multiplying both terms in \eqref{eq:opt1} by $\boldU\ctranspose (f_k)$.

 \begin{figure}[!t]
\centering 
\centering\includegraphics[width=0.7\columnwidth]{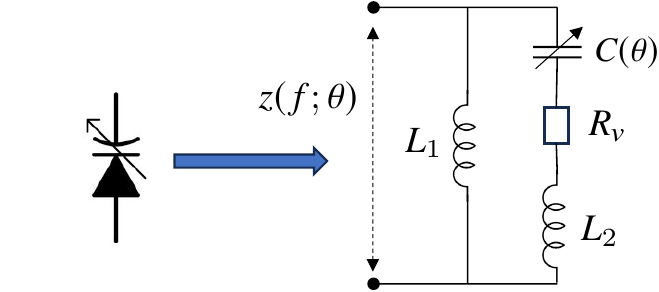}
\caption{Equivalent circuit model of a varactor diode used as reconfigurable load.} 
\label{Fig:VaractorModel}
\end{figure}

\section{Numerical Results}
\label{Sec:NumericalResults}

In this section, we present some numerical results related to the use cases described in  Sec.~\ref{Sec:Optimization} with the purpose of validating and investigating the potential of the  \ac{DSA} with respect to classical array structures and \ac{SIM}. 
The following parameters have been considered in the numerical evaluations if not otherwise specified: $f_0=2.4\,$GHz, $\Gr=0\,$dB, $R=50\,$Ohm,  half-wavelength vertically oriented dipoles. 
 Matrix $\bZ$ has been computed analytically using the coupling model in \cite{BalB:16}. 
 Different and more complex implementations of antenna elements can be considered as well. In such cases, $\bZ$ can be computed off-line using \ac{EM} tools. In \cite{Dar:J24b}, examples using Hertzian dipoles are provided.

\begin{figure}[!t]
\centering 
\centering\includegraphics[width=0.9\columnwidth]{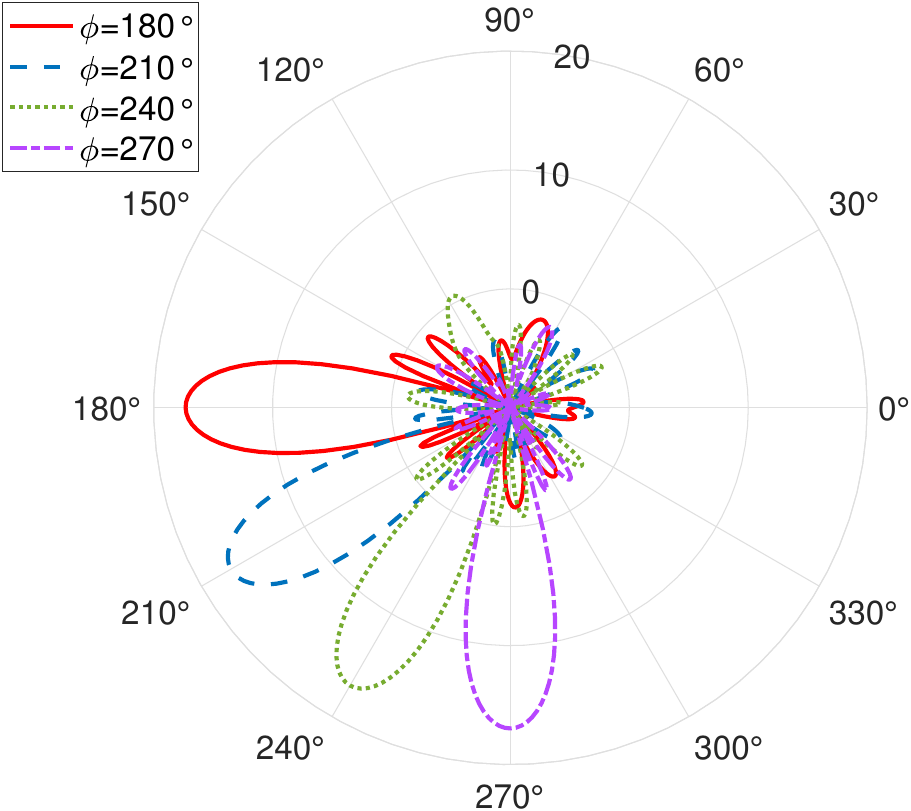}
\caption{Radiation diagram of a disk-shaped DSA ($L_{\text{R}}=1$) with $\Na=1$, $\Delta_{\text{R}}=\lambda/4$, and $L=5$.} 
\label{Fig:PatternDSA}
\end{figure}


\begin{figure}[t]
\centering 
\centering\includegraphics[width=0.95\columnwidth]{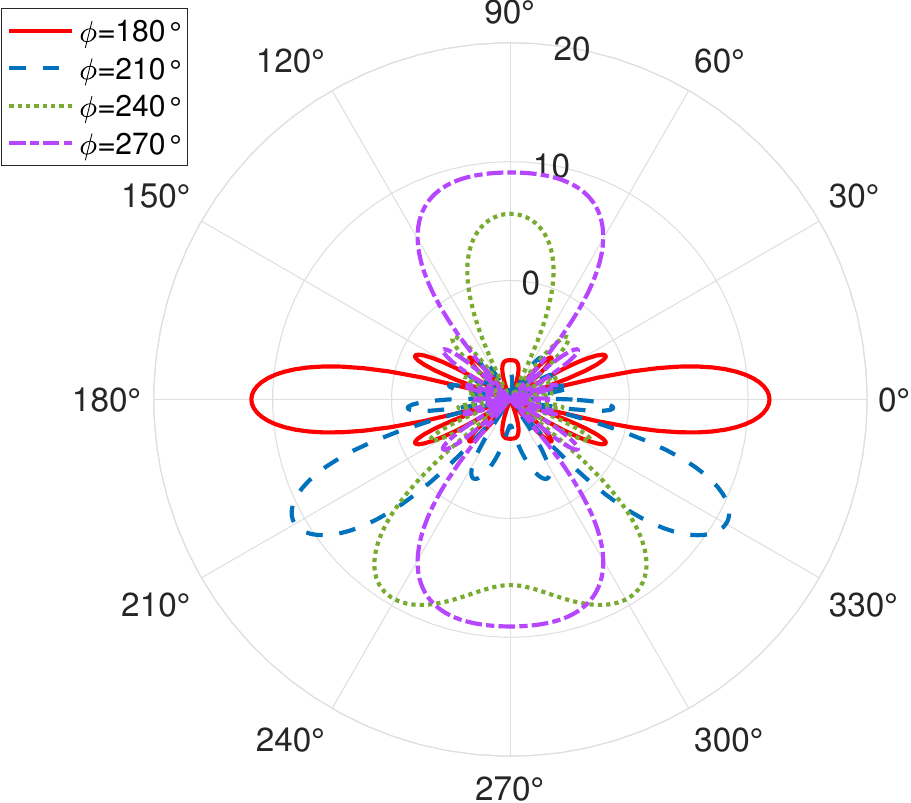}
\caption{Radiation pattern of a conventional ULA with $\Na=7$ active elements.} 
\label{Fig:PatternULA}
\end{figure}

\begin{figure}[t]
\centering 
\centering\includegraphics[width=0.9\columnwidth]{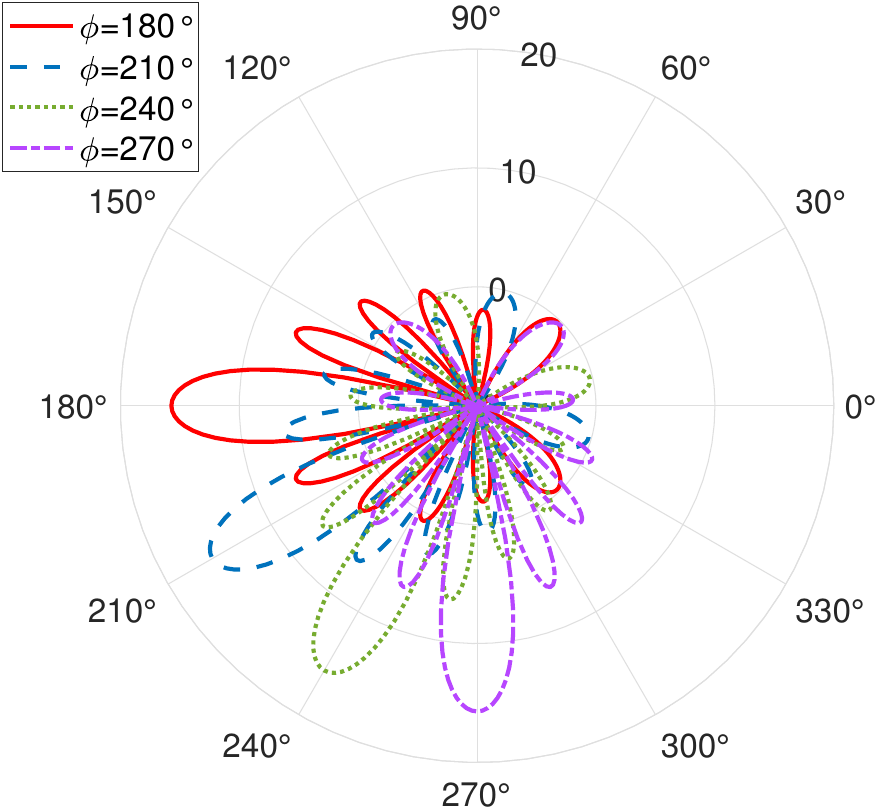}
\caption{Radiation pattern of a conventional UCA with $\Na=36$ active elements.} 
\label{Fig:PatternUCA}
\end{figure}

\subsection{Model for the Reconfigurable Loads}
\label{Sec:Varactor}

Regarding the reconfigurable load, we consider a varactor diode whose equivalent circuit model is reported in Fig. \ref{Fig:VaractorModel},  as suggested in \cite{SamRuiQinCha:20}. By varying the polarization current of the diode (typically a few nA), its equivalent capacitance, and hence its impedance, can be changed.
The corresponding impedance is
\begin{equation}
z(f; \theta)=\frac{j 2 \pi f L_1 (j 2 \pi f  L_2+1/(j 2 \pi f \, C(\theta))+R_v)}{j 2 \pi f (L_1+L_2)+1/(j 2 \pi f \, C(\theta))+R_v} 
\end{equation}
where $C(\theta)=C_{\text{min}}+(C_{\text{max}}-C_{\text{min}}) (\tan^{-1}(\theta)+\pi/2)/\pi$, being $C_{\text{min}}$ and $C_{\text{max}}$, respectively, the minimum and maximum capacitance of the varactor. The relationship $C(\theta)$ between the optimization parameters $\theta$ and $C$ is arbitrary but must be monotonic, and has been chosen so that $\theta$ is unbounded and the optimization problem in \eqref{eq:opt1} remains unconstrained, thus making its numerical solution computation faster.   Varactor diodes exhibit minimal power consumption due to their negligible reverse leakage current so the power consumption of the \ac{DSA} is determined by the digital control circuitry and the specific technology to realize it   \cite{RaiJalTukZhaAbbImr:23}.
In the numerical results, the following parameters have been considered: $R_v=0.1\,$Ohm, $L_1=2.5\,$nH, $L_2=0.7 \,$nH, $C_{\text{min}}=0.47\,$pF, $C_{\text{max}}=2.35\,$pF \cite{SamRuiQinCha:20}.

\subsection{Superdirective Beam Forming}
We first investigate the single-frequency ($K=1$) beam forming capabilities of the \ac{DSA} using one RF chain ($\Na=1$ active antenna element) and the deployment of the surrounding reconfigurable scatterers according to $L_{\text{R}}$ vertical disks (along the $z$-axis) each of them composed of $L$ concentric rings (layers) with incremental radius of step  $\Delta_{\text{R}}\,$ 
(see Fig.~\ref{Fig:Example1}). 
This cylindrical structure appears particularly appealing, especially for its use in base stations or access points, thanks to its circular symmetry. When $L_{\text{R}}=1$, the cylinder degenerates into a disk.  Obviously, other structures can be considered as well, depending on the specific application. 

The target end-to-end channel matrix $\bHopt_1=\bHopt$ with $T=120$ test points deployed on the $x-y$ plane at distance $d=100\,$m (far field) according to the use case in Sec.~\ref{Sec:Example1} was considered for 4 different steering angles, $180^{\circ}$, $210^{\circ}$, $240^{\circ}$, and $270^{\circ}$, respectively. Only the optimization Step 2 was performed with $N_{\text{i}}=1500$ iterations and setting $\Wd{1}=1$ (in this case, $\Wd{1}$ is a scalar and no digital processing takes place).

In Fig.~\ref{Fig:PatternDSA}, the radiation diagrams for the 4 steering angles obtained with $\Delta_{\text{R}}=\lambda/4$, $L=5$, $L_{\text{R}}=1$ (disk shape), corresponding to $\Ns=121$ scatterers are shown. Perfect matching is considered.    
As can be observed, the directivity $D$ of the \ac{DSA} is independent of the angle and it is about $18.2\,$dB. In addition, limited back radiation is obtained without the need to insert a ground plane that would impede the steering in the angle range $[90^{\circ} - 270^{\circ}]$. 
 
 For comparison, we consider the same structure organized as a \ac{SIM} with an impedance matrix having the structure as in \eqref{eq:ZSIM}. The achieved directivity is about $14\,$dB in all directions (not displayed in the figure), corresponding to a performance loss of about $4\,$dB with respect to the \ac{DSA}.
As pointed out in the Introduction, the performance of any \ac{SIM} is constrained by the final layer of the structure, therefore it is upper bounded by that one can obtain by replacing the \ac{SIM} with a fully digital architecture offering unrestricted processing flexibility folowed by a planar or linear array with the same layout and number of elements of the final layer of the \ac{SIM}.   
Therefore, for the sake of comparison, we consider the radiation diagrams obtained using a standard ideal full-digital \ac{ULA} and a \ac{UCA}  with the same aperture of the \ac{DSA} ($36\,$cm diameter) in the $x-y$ plane. The corresponding radiation diagrams are reported in Figs.~\ref{Fig:PatternULA} and \ref{Fig:PatternUCA}, respectively. These arrays require $\Na=7$ and $\Na=36$ active antennas compared to only one active antenna of the \ac{DSA}.
As well-known, in \acp{ULA}  the directivity degrades when approaching $90^{\circ}$ from a maximum value of $11.7\,$dB (about $10\log_{10}(\Na \, G_{\text{d}})$, with $G_{\text{d}}$ the directivity of the half-wave dipole) and symmetric back radiation is present. Instead, with the \ac{UCA}, the directivity ($15.7\,$dB) is insensitive to the steering direction at the expense of higher side lobes. Nevertheless, in both cases, the proposed \ac{DSA} exhibits superdirectivity capability with an additional gain of $6.5\,$dB and $2.5\,$dB, respectively. Contrary to standard arrays, where superdirectivity is obtained only in the end-fire direction \cite{IvrNos:14}, here notably, superdirectivity is equally obtained in all steering directions.  
Superdirectivity cannot be obtained instead using a \ac{SIM}-like structure.

\begin{table}[tp]
\caption{$L_{\text{R}}=1$, $L=5$.}
\begin{center}
\begin{tabular}{|c|c||c|c|c||c|c|c|c|}
\hline
\multicolumn{2}{|c||}{Configuration}  & 
\multicolumn{3}{|c||}{Perfect matching}  & 
\multicolumn{4}{|c|}{Simplified matching} \\
\hline
 $\Delta_{\text{R}}$ & $\Ns$ & $D\,$(dB) & $\etad$ & $Q$ & $D\,$(dB) & $\etam$ & $\etad$ & $Q$ \\
 \hline
$\lambda/8$&   89& 16.4 & 0.78 &19.3 & 16 & 1 & 0.81  & 16.6 \\
$\lambda/6$ &   100 &17  & 0.8 & 19.4 & 17  &  1 & 0.82 & 15.7 \\ 
$\lambda/4$  &   121 & 18.2 & 0.79 & 19.6 & 18.2 & 1 & 0.79 & 16.8\\
$\lambda/2$ &  184 &15.6 & 0.82 & 4.82 & 15.7 &  0.92 & 0.92 & 4.3 \\
$\lambda$ &   310 & 15.8 & 0.93  & 4.58 & 15 & 0.86 & 0.96 & 1.03\\ 
\hline
\end{tabular}
\end{center}
\label{Table:Cylinder1}
\end{table}

\begin{table}[tp]
\caption{$L_{\text{R}}=1$, $\Delta_{\text{R}}=\lambda/4$.}
\begin{center}
\begin{tabular}{|c|c|c|c|c|c|}
\hline
 Configuration &  $\Ns$ & $D\,$(dB) & $\etam$ & $\etad$ & $Q$-factor \\
 \hline
$L=3$ & 53 & 15.4 & 1 & 0.84 & 12.9 \\
$L=4$ & 84 & 16.4 & 1 & 0.82 & 15.7 \\ 
$L=5$ & 121  & 18.2  & 1 & 0.79 &  16.8 \\ 
$L=6$ & 174 & 19.1 & 1 & 0.78 & 17.6 \\
\hline
\end{tabular}
\end{center}
\label{Table:Cylinder2}
\end{table}

\begin{table}[tp]
\caption{$L=5$, $\Delta_{\text{R}}=\lambda/4$.}
\begin{center}
\begin{tabular}{|c|c|c|c|c|c|}
\hline
 Configuration & $D\,$(dB) & $\Ns$ & $\etam$ & $\etad$ & $Q$-factor \\
 \hline
$L_{\text{R}}=1$ &121  & 18.2 & 1 & 0.79 & 16.8  \\
$L_{\text{R}}=2$ & 168 & 18.4 & 1 & 0.82 & 13.5 \\ 
$L_{\text{R}}=3$ & 252 & 19.4 & 0.98 & 0.79 & 14.3 \\ 
\hline
\end{tabular}
\end{center}
\label{Table:Cylinder3}
\end{table}


\begin{table}[tp]
\caption{Random shape. $L_{\text{R}}=1$}
\begin{center}
\begin{tabular}{|c|c|c|c|c|}
\hline
$\Ns$ & 60 & 121 & 242 & 400  \\ 
\hline
$D\,$(dB) & 16 & 17.8  & 19  & 19.5  \\
$\eta_m$ & 0.97 & 1 & 1 & \\
$\eta_d$ & 0.85 & 0.78 & 0.73& 0.66 \\
 $Q$-factor & 10.71 & 16.5  & 21  & 26.3  \\
 \hline
\end{tabular}
\end{center}
\label{Table:Random}
\end{table}

In Tables~\ref{Table:Cylinder1}-\ref{Table:Cylinder3}, the impact of the layers spacing $\Delta_{\text{R}}$, number of layers $L$, and the number of disks $L_{\text{R}}$ is investigated. 
As it can be noticed in Table~\ref{Table:Cylinder1}, the maximum directivity is obtained for  $\Delta_{\text{R}}=\lambda/4$, confirming that superdirectivity and a higher $Q$ factor can be achieved only by exploiting the mutual coupling between the antenna elements but, at the same time, that there is an optimum spacing value to be found. 
In fact, on one hand, higher spacing reduces the coupling and hence the contribution to radiation of those scatterers located far away from the active element. On the other hand, very close spacing tightens the coupling between the elements and reduces the \ac{DoF} available for optimization. 
The impact of the dissipative elements on the efficiency $\etad$ is more evident with close spacing between the elements (higher mutual coupling). 

In Table~\ref{Table:Cylinder1}, the results for a \ac{DSA} employing both the perfectly matched network and the simplified one proposed in Sec. \ref{Sec:SimpleMatching} are reported.  The comparison highlights a negligible difference in terms of directivity and that an almost perfect matching is reached in most of the configurations ($\etam \approx 1$). Therefore, in the following results, only the simplified matching network is considered. 
A higher number of layers $L$ (3D cylindric structure) increases the directivity in the $x-y$ plane at the expense of a higher number of reconfigurable scattering elements $\Ns$, as reported in Table \ref{Table:Cylinder2}. 
The directivity can be further increased by thinning the radiation diagram in the $z$ direction by introducing more disks $L_{\text{R}}$, i.e., moving from a dish shape to a cylindrical shape, and it can be observed in Table \ref{Table:Cylinder3}.  

An example using a random deployment of the scatterers within a disk as a function of their density is provided in Table~\ref{Table:Random} for a fixed diameter of $36\,$cm. As can be seen, the achievable performance is similar to that of the regular cylindrical deployment when compared with the same number of scatterers, even though, in general, higher values of $Q$-factor are obtained because of the possibility that a couple of scatterers are very close to each other.
The analysis and optimization of the deployment strategies of the scatterers is a topic that deserves future investigations.     

An interesting aspect worth consideration is the sensitivity of the performance to parameter mismatch that might arise in practical implementations. We evaluated the directivity degradation for the configuration in Fig.~\ref{Fig:PatternDSA} when the actual implemented $n$th parameter of the \ac{DSA}, $\hat{\theta}_n$, is affected by an error $d_n$ modeled as a zero-mean  Gaussian random variable with standard deviation $\sigma_n$, i.e., $\hat{\theta}_n=\theta_n+d_n$. The corresponding directivity obtained for  $\sigma_n=0$, $\sigma_n=10\%$ and $\sigma_n=20\%$ of the nominal absolute value $|\theta_n|$ are, respectively, $D=18.2\,$dB ($\etam \simeq 1$), $D=17.7\,$dB ($\etam=0.86$), and $D=15.5\,$dB ($\etam=0.85$), showcasing about $3\,$dB loss with $20\%$ relative error.  

\begin{figure}[t]
\centering 
\centering\includegraphics[width=0.9\columnwidth]{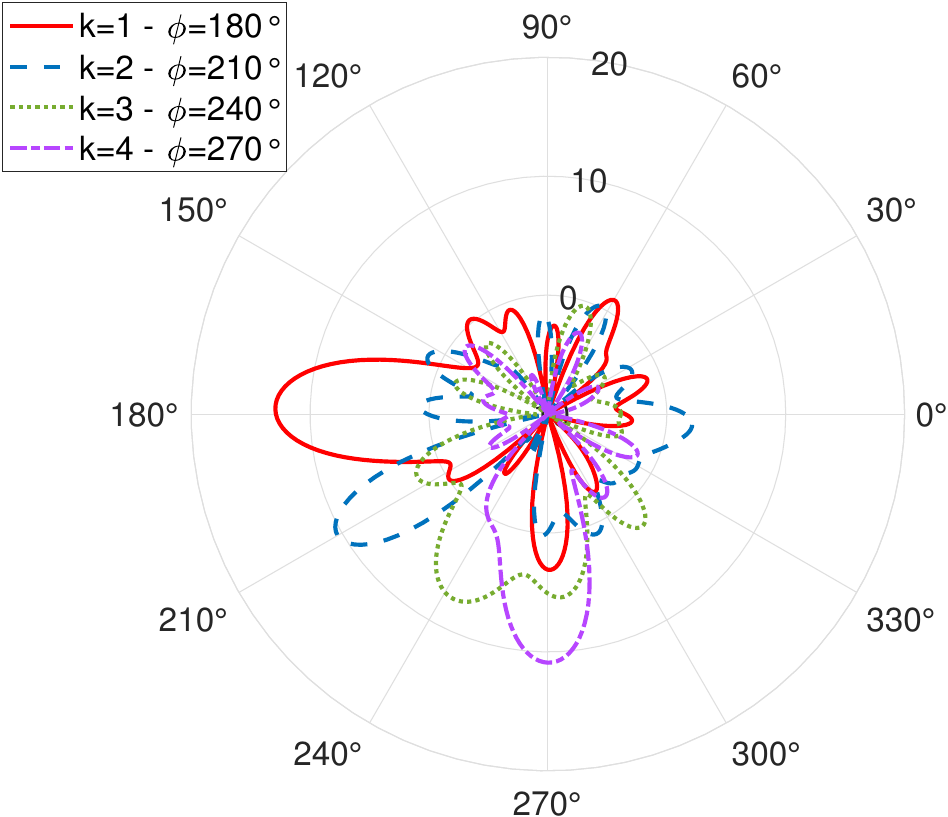}
\caption{Multi-frequency radiation diagrams of a disk-shaped DSA with $\Delta_{\text{R}}=\lambda/4$, $L=5$, $\Na=1$, $K=4$ subcarriers, and $W=80\,$MHz.} 
\label{Fig:RadiationSingleFreq}
\end{figure}

The capability of the \ac{DSA} to provide frequency-selective radiation diagrams can be observed in Fig.~\ref{Fig:RadiationSingleFreq}, obtained with one \ac{RF} chain and   $K=4$ subcarriers distributed within the bandwidth $W=80\,$MHz. The target steering angles for each subcarrier are: $180^{\circ}$, $210^{\circ}$, $300^{\circ}$, and $330^{\circ}$. Compared to the single-carrier situation, the diagrams present slightly lower directivities and higher sidelobes, indicating that managing wideband signals with the same configuration parameters of the \ac{DSA} is more challenging. The motivation might be ascribed to the usage of dipoles with length equal to the wavelength at the carrier frequency $f_0$, whereas the actual used frequency differs from $f_0$ for $k>1$.

\begin{figure}[t]
\centering 
\centering\includegraphics[width=0.9\columnwidth]{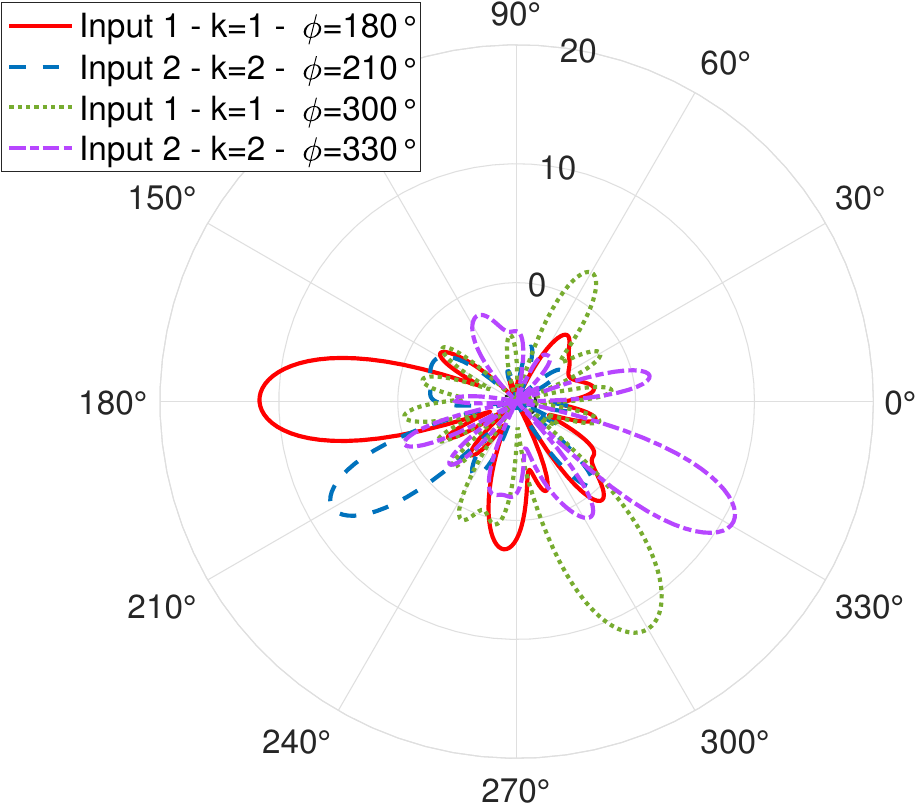}
\caption{Multi-frequency multi-input radiation diagrams of a disk-shaped DSA with $\Delta_{\text{R}}=\lambda/4$, $L=5$, $\Na=2$, $K=2$ subcarriers, and $W=80\,$MHz.} 
\label{Fig:RadiationNa2Nc2}
\end{figure}

An example of multi-functional and multi-frequency design is reported in Fig.~\ref{Fig:RadiationNa2Nc2}, where each \ac{RF} chain ($\Na=2$ in this case) is associated with a different response for $K=2$ different frequencies. The target steering angles for each input/subcarrier (sc) combination are: $180^{\circ}$ (input 1, sc 1), $210^{\circ}$ (input 1, sc 2), $300^{\circ}$ (input 2, sc 1), and $330^{\circ}$ (input 2, sc 2). The plots show that with the same configuration $\btheta$, obtained from the minimization of \eqref{eq:opt}, the \ac{DSA} is capable of associating a dedicated beam with each input/subcarrier combination. 
While the resulting diagrams demonstrate the interesting flexibility of \ac{DSA} in providing flexible frequency-selective responses, the reduced gain indicates that the wideband design of \ac{DSA} deserves further investigation, left to future works.

\subsection{Multi-user \ac{MISO}}
We consider now the scenario of use case 2 in Sec.~\ref{Sec:Example2} illustrated in Fig.~\ref{Fig:Example2}, where a \ac{DSA} with $\Delta_{\text{R}}=\lambda/4$, $L_{\text{R}}=1$,  $\Na=4$ active antennas and $\Na$ RF chains is used to communicate simultaneously with $T=\Na$ single-antenna users located at a distance of $d=100\,$m and angles $-40^{\circ}$, $20^{\circ}$, $80^{\circ}$, and $200^{\circ}$ in the $x-y$ plane.  The \ac{DSA} has been optimized according to \eqref{eq:ZF} (zero forcing criterium) with $N_{\text{i}}=50$ and $N_{\text{alt}}=10$. 

\begin{figure}[!t]
\centering 
\centering\includegraphics[width=1\columnwidth]{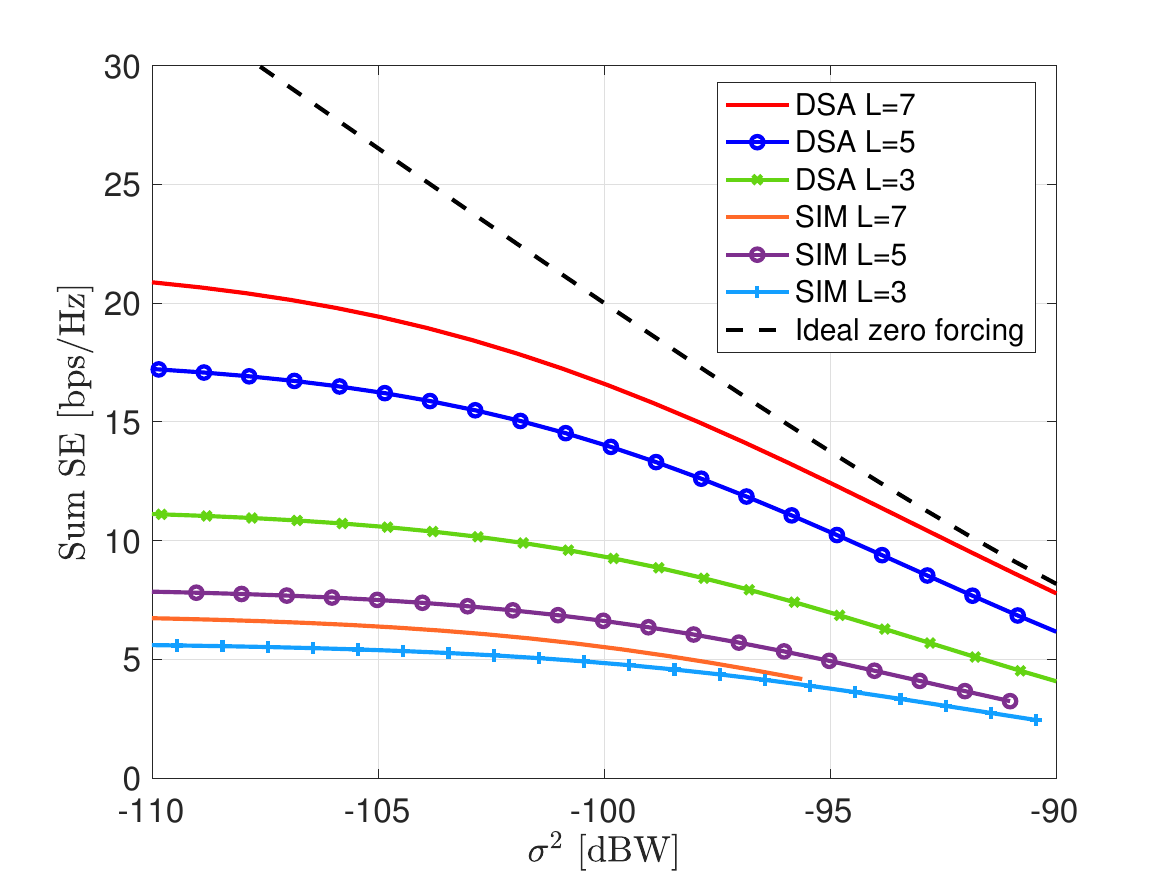}
\caption{Sum spectral efficiency of a zero-forcing 4-user MISO system based on a DSA and a SIM for different numbers of layers $L$.} 
\label{Fig:MISO}
\end{figure}

The corresponding performance is shown in Fig.~\ref{Fig:MISO}, where the sum \ac{SE} is plotted as a function of the power $\sigma^2$ of the \ac{AWGN} for a transmitted power $\Ptx=10\,$mW and for different numbers of layers $L$. 
The results obtained with a \ac{SIM} configuration employing the same geometry as well as with an ideal system with perfect zero forcing and directivity of $D=17\,$dB
are also reported for comparison. 
The presence of a saturation for small $\sigma^2$, i.e., higher \acp{SNR}, reveals an imperfect nulling of the multi-user interference. The \ac{DSA} showcases much higher sum \ac{SE} corresponding to a better approximation of the zero forcing criterion than the \ac{SIM} for the same geometry. In this setup, the \ac{DSA} provides the best performance with $L=7$ layers, whereas the \ac{SIM} exhibits a tradeoff whose optimum is $L=5$ layers.

\subsection{MIMO EM Precoder}
Finally, we consider the scenario of use case 3 in Sec.~\ref{Sec:Example3} illustrated in Fig.~\ref{Fig:Example3}, where a multi-layer \ac{MIMO} link with a user located at $d=10\,$m equipped with a standard \ac{ULA} with $20$ elements spaced of $\lambda/2$ has to be established in \ac{NLOS} condition.
The \ac{DSA} is equipped with $\Na=4$ active antennas and $\Na$ RF chains to allow the transmission of up to 4 data streams (layers) according to channel characteristics.
The simulated \ac{NLOS} channel consists of 5 multi-paths caused by the reflection of 5 scatterers located at angles $\phi=\{-43^{\circ},-14^{\circ},14^{\circ}, 43^{\circ}, 72^{\circ} \}$ and distance $5\,$m from the \ac{DSA}. 
The corresponding strongest singular values of the channel are $\bLambda=\diag{-37.3{\,\text{dB}},-39.8{\,\text{dB}},-42{\,\text{dB}},-48.8{\,\text{dB}}, -82{\,\text{dB}}, \ldots }\,$, which has clearly rank $r=4$.
The \ac{DSA} has the following parameters $\Delta_{\text{R}}=\lambda/4$, $L_{\text{R}}=1$, $\Ns=121$, and it is optimized according to the criterium in \eqref{eq:SVD} with $N_{\text{i}}=50$ and $N_{\text{alt}}=10$.

In Table~\ref{Table:Lambda}, the resulting coefficients $\hat{\bLambda}$ of the end-to-end channel are reported. The comparison between the diagonal and off-diagonal values indicates that the coupling between different layers is completely negligible, i.e., the channel is almost perfectly diagonalized. Compared to the singular values of the channel in $\bLambda$, the diagonal elements of $\hat{\bLambda}$ exhibit the same behavior with a loss of about $7-8\,$dB that can be ascribed to the fact that the \ac{DSA} is not ideal.  
The effect of implementation errors is reported in the bottom part of the table for $1\%$ and $10\%$ relative error standard deviation, respectively. It can be seen that the effect of implementation errors becomes no longer negligible (coupling above $-10\,$dB) with $10\%$ relative error.   
In addition, numerical investigations revealed that the presence of the digital processor is not fundamental from the performance viewpoint, but slightly speeds up the convergence of the optimization process.

\section{Conclusion}
\label{Sec:Conclusion}

In this paper, we have presented a general framework for the characterization and optimization of frequency-selective \acfp{DSA}.
This has been achieved through an accurate modeling of the coupling between the scattering elements and their loads, and the joint optimization of \ac{EM} processing, power matching,   and radiation of the active and reconfigurable scattering elements interacting in the reactive near field.
It has been shown that a \ac{SIM} can be modeled as a particular case of a \ac{DSA} and that, for a given geometry, the \ac{DSA} outperforms \ac{SIM} in approximating the desired processing and enables the exploitation of superdirective capabilities.

The results obtained in this paper demonstrate the potential of \ac{DSA} to realize multifunctional and multi-frequency processing tasks ``over the air" by leveraging the coupling between active and scattering elements, thus minimizing the number of \ac{RF} chains and baseband processing in the digital domain. This is achieved through a semi-passive structure with minimal energy consumption and nearly zero latency, as the processing occurs at the speed of light. 

\begin{table}[tp]
\caption{${\left [ \hat{\Lambda} \right ]  }_{n, k}$ in dB, for $n, k=1, 2, 3, 4$.}
\begin{center}
\begin{tabular}{|c|c|c|c|}
\hline
\multicolumn{4}{|c|}{Perfect implementation }    \\
\hline
-45.2  & -184  & -186 & -175 \\
 -175 & -48 & -190 & -177\\
 -182 & -183 &  -49.8 & -174 \\
 -172 & -174 & -165  & -56.7 \\
 \hline
 \hline
\multicolumn{4}{|c|}{$1\%$ relative implementation error}    \\
\hline
-45.6 &  -67  & -81  &  -74 \\
-69  & -47.3  & -86  & -75 \\ 
-68  & -78  & -49.6  & -71 \\
-75 & -84  & -99  & -57  \\
\hline
 \hline
\multicolumn{4}{|c|}{$10\%$ relative implementation error}  \\
\hline
-43.5  & -57  & -56  & -59 \\
  -66.6  & -48  & -55  & -60 \\
  -68  & -67  & -51  & -69 \\
  -71  & -66  & -70 & -54 \\
\hline
\end{tabular}
\end{center}
\label{Table:Lambda}
\end{table}


\section*{Acknowledgment}
This work was supported by the European Union under the Italian National Recovery and Resilience Plan (NRRP) of NextGeneration EU, partnership on ``Telecommunications of the Future" (PE00000001 - program ``RESTART"), by the HORIZON-JU-SNS-2022-STREAM-B-01-03 6G-SHINE project (Grant Agreement No. 101095738), and by the HORIZON-JU-SNS-2022 project TIMES (Grant no. 101096307).

\ifCLASSOPTIONcaptionsoff
\fi
\bibliographystyle{IEEEtran}

\bibliography{IEEEabrv,Biblio/BiblioDD,Biblio/MetaSurfaces,Biblio/EMInformationTheory,Biblio/IntelligentSurfaces,Biblio/MassiveMIMO,Biblio/MIMO,Biblio/THzComm,Biblio/EMTheory,Biblio/WINS-Books,Biblio/Vari}

\end{document}